# Synaptic plasticity as Bayesian inference

Laurence Aitchison[*,1,2], Jannes Jegminat[3,4], Jorge Aurelio Menendez[1,5], Jean-Pascal Pfister[3,4], Alex Pouget[1,6] and Peter E. Latham[1]

[*]Corresponding author

[1]Gatsby Computational Neuroscience Unit, UCL, London, UK

[2]University of Bristol, Bristol, UK

[3]Institute of Neuroinformatics, UZH/ETH Zurich, Zurich, Switzerland

[4]Department of Physiology, University of Bern, Bern, Switzerland

[5]CoMPLEX, UCL, London, UK

[6]University of Geneva, Geneva, Switzerland

**Abstract**

Learning, especially rapid learning, is critical for survival. However, learning is hard: a large number of synaptic weights must be set based on noisy, often ambiguous, sensory information. In such a high-noise regime, keeping track of probability distributions over weights is the optimal strategy. Here we hypothesize that synapses take that strategy; in essence, when they estimate weights, they include error bars. They then use that uncertainty to adjust their learning rates, with more uncertain weights having higher learning rates. We also make a second, independent, hypothesis: synapses communicate their uncertainty by linking it to variability in PSP size, with more uncertainty leading to more variability. These two hypotheses cast synaptic plasticity as a problem of Bayesian inference, and thus provide a normative view of learning. They generalize known learning rules, offer an explanation for the large variability in the size of post-synaptic potentials, and make falsifiable experimental predictions.

## 1 Introduction

To survive, animals must accurately estimate the state of the world. This estimation problem is plagued by uncertainty: not only is information often extremely limited (e.g., because it is dark) or ambiguous (e.g., a rustle in the bushes could be the wind, or it could be a predator), but sensory receptors, and indeed all neural circuits, are noisy. Historically, models of neural computation ignored this uncertainty, and relied instead on the idea that the nervous system estimates values of quantities in the world, but does not include error bars [1]. However, this does not seem to be what animals do — not only does ignoring uncertainty lead to suboptimal decisions, it is inconsistent with a large body of experimental work [2, 3]. Thus, the current view is that in many, if not most, cases, animals keep track of uncertainty, and use it to guide their decisions [3].

Accurately estimating the state of the world is just one problem faced by animals. They also need to learn, and in particular they need to leverage their past experience. It is believed that learning primarily involves changing synaptic weights. But estimating the correct weights, like estimating the state of the world, is plagued by uncertainty: not only is the information available to synapses often extremely limited (in many cases just pre and post synaptic activity), but that information is highly unreliable. Historically, models of synaptic plasticity ignored this uncertainty, and assumed that synapses do not include error bars when they estimate their weights. However, uncertainty is important for optimal learning — just as it is important for optimal inference of the state of the world.

Motivated by these observations, we propose two hypotheses. The first, Bayesian Plasticity (so named because it is derived using Bayes' rule), states that during learning, synapses do indeed take uncertainty into account. Under this hypothesis, synapses do not just estimate what their weight should be, they also include error bars. This allows synapses to adjust their learning rates on the fly: when uncertainty is high learning rates are turned up, and when uncertainty is low learning rates are turned down. We show that these adjustments allow synapses to learn faster, so there is likely to be considerable evolutionary pressure for such a mechanism. And indeed, the same principle has recently been shown to recover state-of-the-art adaptive optimization algorithms for artificial neural networks [4].

Bayesian Plasticity is a hypothesis about what synapses compute. It does not, however, tell synapses how to set their weights. For that a second hypothesis is needed. Here we propose that weights are sampled from the probability distribution describing the synapse's degree of uncertainty. Under this hypothesis, which we refer to as Synaptic Sampling, trial to trial variability provides a readout of uncertainty: the larger the trial to trial variability in synaptic strength, the larger the uncertainty. Synaptic Sampling is motivated by the observation that the uncertainty associated with a particular computation should depend on the uncertainty in the weights. Thus, to make optimal decisions, the brain needs to know something about the uncertainty; one way for synapses to communicate that is via variability in the postsynaptic potential (PSP) amplitude (see Supplementary Math Note, Sec. S7, for an extended discussion).

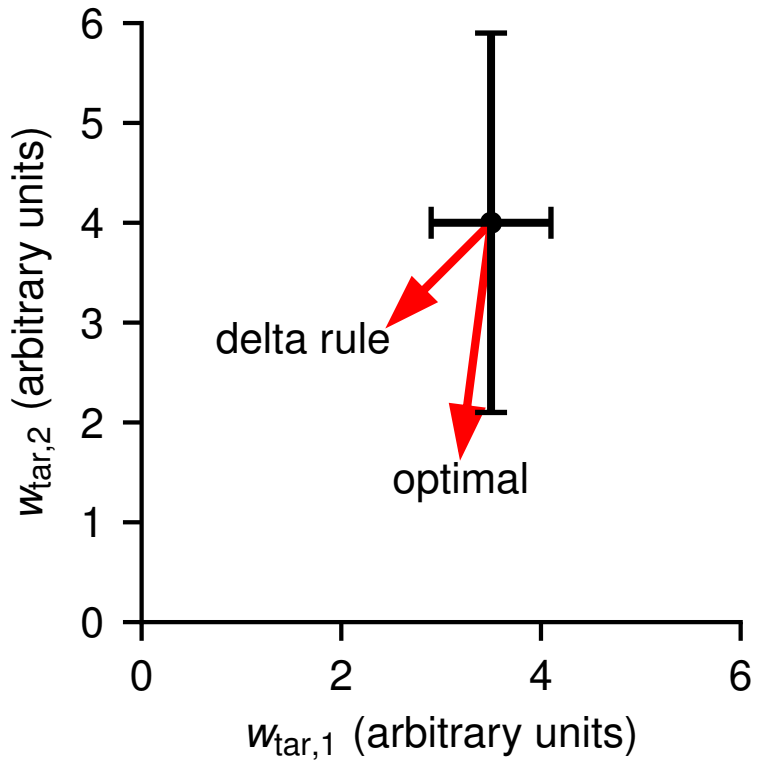

Figure 1: The delta rule is suboptimal. The error bars denote uncertainty (measured by the standard deviation around the mean) in two synapses' estimates of their target weights, $w_{\text{tar},1}$ and $w_{\text{tar},2}$. The first is reasonably certain; the second less so. The red arrows denote possible changes in response to a negative feedback signal. The arrow labeled "delta rule" represents an equal decrease in the first and second target weights. In contrast, the arrow labeled "optimal" takes uncertainty into account, so there is a larger change in the second, more uncertain, target weight.

Combined, these two hypotheses make several strong experimental predictions. As we discuss below, one is consistent with re-analysis of existing experimental data; the others, which are feasible in the not so distant future, could falsify one or both hypotheses. We begin by analyzing the first hypothesis, that synapses keep track of their uncertainty (Bayesian Plasticity); after that we discuss our second hypothesis, that synapses sample from the resulting distribution (Synaptic Sampling).

# 2 Results

Under Bayesian Plasticity, each synapse computes its mean and variance, and updates both based on the pattern of presynaptic spikes. In analogy to classical learning rules, the update rule for the mean pushes it in a direction that reduces a cost function. But in contrast to classical learning rules, the amount the mean changes depends on the uncertainty: the higher the uncertainty, as measured by the variance, the larger the change in the mean. The variance thus sets the learning rate, as shown in Fig. 1. In essence, there is a rule for computing the learning rate of each synapse.

To illustrate these ideas, we consider a model of synaptic integration in which postsynaptic potentials combine linearly,

$$V(t) = \sum_i w_i(t) x_i(t) + \xi_V(t) \tag{1}$$

where $V(t)$ is the membrane potential, $x_i(t)$ is the synaptic input from neuron $i$, $w_i(t)$ is the corresponding PSP amplitude, and $\xi_V(t)$ is the membrane potential noise. For simplicity we work in discrete time, so $x_i(t)$ is either 1 (when there is a spike at time $t$) or 0 (when there is no spike), and we take the time step to be 10 ms, on the order of the membrane time constant [5].

We assume that the goal of the neuron is to set its weights, $w_i$, so that it achieves a "target" membrane potential (denoted $V_{\text{tar}}$) – the membrane potential that minimizes some cost to the animal. In this setting, the weights are found using a neuron-specific feedback signal, denoted $f$. Critically, this feedback signal contains information about the target weights through its dependence on a true error signal, $\delta$ – the difference between the target and actual membrane potential,

$$\delta \equiv V_{\text{tar}} - V. \tag{2}$$

Our focus is on how to use the feedback signal most efficiently, not on where it comes from (an active area of research [6, 7, 8, 9, 10]). Thus, in most of our analysis we simply assume that the neuron receives a feedback signal, and ask how to optimally update the weights via Bayesian inference.

We consider several learning scenarios. In the first, we simply add noise, denoted $\xi_\delta$, to $\delta$, resulting in the error signal $f_{\text{lin}} = \delta + \xi_\delta$ (the subscript "lin" indicates that the average feedback is linear in $\delta$). The second corresponds to cerebellar learning, in which a Purkinje cell receives a complex spike if its output is too high, thus triggering long term depression [11]. To mimic the all-or-nothing nature of a complex spike [12], we use a cerebellar-like feedback signal: $f_{\text{cb}} = \Theta(\delta + \xi_\delta - \theta)$ where $\Theta$ is the Heaviside step function. For this feedback signal, $f_{\text{cb}}$ is likely to be 1 if $\delta$ is above a threshold, $\theta$, and likely to be 0 if it is below threshold. The third corresponds to reinforcement learning, in which the feedback represents the reward. The reward provides the magnitude of the error signal, but not its sign, so the feedback signal is $f_{\text{rl}} = -|\delta + \xi_\delta|$. In the fourth scenario we move beyond analysis of single neurons, and consider learning the output weights of a recurrent neural network. In this scenario, the error signal is $\delta$, without added noise.

The main idea behind Bayesian plasticity is most easily illustrated in the simplest possible setting, linear feedback, $f_{\text{lin}} = \delta + \xi_\delta$. In that case there is a well known learning rule, the delta rule [13, 14],

$$\Delta w_i = \eta x_i f_{\text{lin}} \,. \tag{3}$$

(This is most easily recognized as the delta rule in the absence of noise, so that $f_{\text{lin}} = \delta$.) The change in the weight is the product of a learning rate, $\eta$, a presynaptic term, $x_i$ and a postsynaptic term, $f_{\text{lin}}$. Importantly, the learning rate, $\eta$, is the same for all synapses, so all synapses whose presynaptic cells are active (i.e., for which $x_i = 1$) change by the same amount (the red arrow labeled "delta rule" in Fig. 1).

In the absence of any other information, the delta rule is perfectly sensible. However, suppose, based on previous information, that synapse 1 is relatively certain about its target weight, whereas synapse 2 is uncertain (error bars in Fig. 1). In that case, new information should have a larger effect on synapse 2 than synapse 1, so synapse 2 should update the estimate of its weight more than synapse 1 (red arrow labeled "optimal" in Fig. 1).

Implementing this scheme leads to several features that are not present in classical learning rules. First, the variance needs to be inferred; second, the change in the weight must depend on the inferred variance; and third, because of uncertainty, the "weight" is in fact the inferred mean weight. In Supplementary Math Note, Sec. S3.3, we derive approximate learning rules that take these features into account (see Methods, Sec. M2, for the exact rules). Using $\mu_i$ and $\sigma_i^2$ to denote the inferred mean and variance of the distribution over weights, those learning rules are

$$\Delta\mu_i \approx \frac{\sigma_i^2}{\sigma_\delta^2}\, x_i\, f_{\text{lin}} - \frac{1}{\tau}\left(\mu_i - \mu_{\text{prior}}\right) \tag{4a}$$

$$\Delta\sigma_i^2 \approx -\frac{\sigma_i^4}{\sigma_\delta^2}\, x_i^2 - \frac{2}{\tau}\left(\sigma_i^2 - \sigma_{\text{prior}}^2\right) \tag{4b}$$

where $\sigma_\delta^2$ is the variance of $f_{\text{lin}}$, and $\tau$, $\mu_{\text{prior}}$, and $\sigma_{\text{prior}}^2$ are fixed parameters (described shortly). Note that $\sigma_i$ corresponds to the length of the error bars in Fig. 1.

The update rule for the mean weight, Eq. (4a), is similar to the delta rule, Eq. (3). There are, however, two important differences. First, the fixed learning rate, $\eta$, that appears in Eq. (3) has been replaced by a variable learning rate, $\sigma_i^2/\sigma_\delta^2$, which is proportional to the synapse's uncertainty, as measured by $\sigma_i^2$. Thus, the more uncertain a synapse is about its target weight, the larger the change in its mean weight when new information arrives — exactly what we expect given Fig. 1. Moreover, as the feedback signal gets noisier (as measured by the variance of $f_{\text{lin}}$), and thus less informative, the learning rate falls. Second, in the absence of information ($x_i = 0$, meaning no spikes), the inferred mean weight, $\mu_i$, moves toward the prior, $\mu_{\text{prior}}$. That is because we are considering the realistic case in which the target weights drift randomly over time due to changes in the statistics of the world and/or surrounding circuits. See Methods, Sec. M1.1 for a detailed discussion of this point.

Unlike the update rule for the mean, the update rule for the uncertainty, $\sigma_i^2$ (Eq. 4b), does not have a counterpart in classical learning rules. It does, however, have a natural interpretation. The first term in Eq. (4b) reduces uncertainty (note the negative sign) whenever the presynaptic cell is active ($x_i = 1$); that is, whenever the synapse receives information. The second term has the opposite effect: it continually increases uncertainty (up to the prior uncertainty, $\sigma_{\text{prior}}^2$), independent of presynaptic spikes. That term arises because random drift slowly reduces knowledge about the target weights.

The learning rules given in Eq. (4) are approximate – their form was optimized for ease of interpretation rather than accuracy. However, the more exact learning rules (Methods, Eqs. (M.41), (M.43) and (M.47), for the three feedback signals) are not that different. In particular, they retain the same flavor: they consist of a presynaptic term ($x_i$) and a postsynaptic term (a function of $f_{\text{lin}}$), and the effective learning rate is updated on each timestep. Moreover, the interpretation is the same: the mean is moved, on average, toward its true value, with a rate that scales with uncertainty, and whenever there is a presynaptic spike the uncertainty is reduced.

To determine whether our Bayesian learning rules are able to accurately compute the mean and variance of the weights, we generated a set of target weights, denoted $w_{\text{tar},i}$, and used those to construct $V_{\text{tar}}$,

$$V_{\text{tar}}(t) = \sum_i w_{\text{tar},i}(t) x_i(t) \,. \tag{5}$$

Simulations show that the mean weights track the target weights very effectively (Fig. 2; compare the black and red lines, which correspond to the target weight and its inferred mean, respectively). Just as importantly, the synapse's estimate of its uncertainty tracks the difference between its estimate and the actual target (the black line should be inside the 95% confidence interval – the red area in Fig. 2 – 95% of the time, and it is very close to that: linear, 96.1%; cerebellar learning, 95.4%; reinforcement learning, 96.8%). Note that the uncertainty is much lower at high presynaptic firing rate than at low rate (the red regions in the top row of Fig. 2 are much narrower than in the bottom row). That is because for low firing rate $x_i$ is mainly zero, and so there is little decrease in $\sigma_i^2$; see Eq. (4b).

The critical aspect of the learning rules in Eq. (4) is that the learning rate — the change in mean PSP amplitude, $\mu_i$, per presynaptic spike — increases as the synapse's uncertainty, $\sigma_i^2$, increases. This is a general feature of our learning rules, and not specific to any one of them. Consequently, independent of the learning scenario, we expect performance to be better than for classical learning rules, which do not take uncertainty into account. To check whether this is true, we computed the mean squared error between the actual and target membrane potential, $V$ and $V_{\text{tar}}$, for classical learning rules, and compared it to the Bayesian learning rules. The results are shown in Fig. 3. The black line in each panel is the mean squared error for the classical learning rules as a function of learning rate; the red line is the mean squared error for the Bayesian learning rules. (The latter is constant because the Bayesian learning rules do not depend on the classical learning rate.) As predicted, the Bayesian learning rules always do better than the classical ones, even if the learning rate is tuned to its optimal value. This result was robust to model mismatch (Supplementary Math Note, Fig. S.11).

For the examples so far, we considered a single neuron inferring only its own input weights.

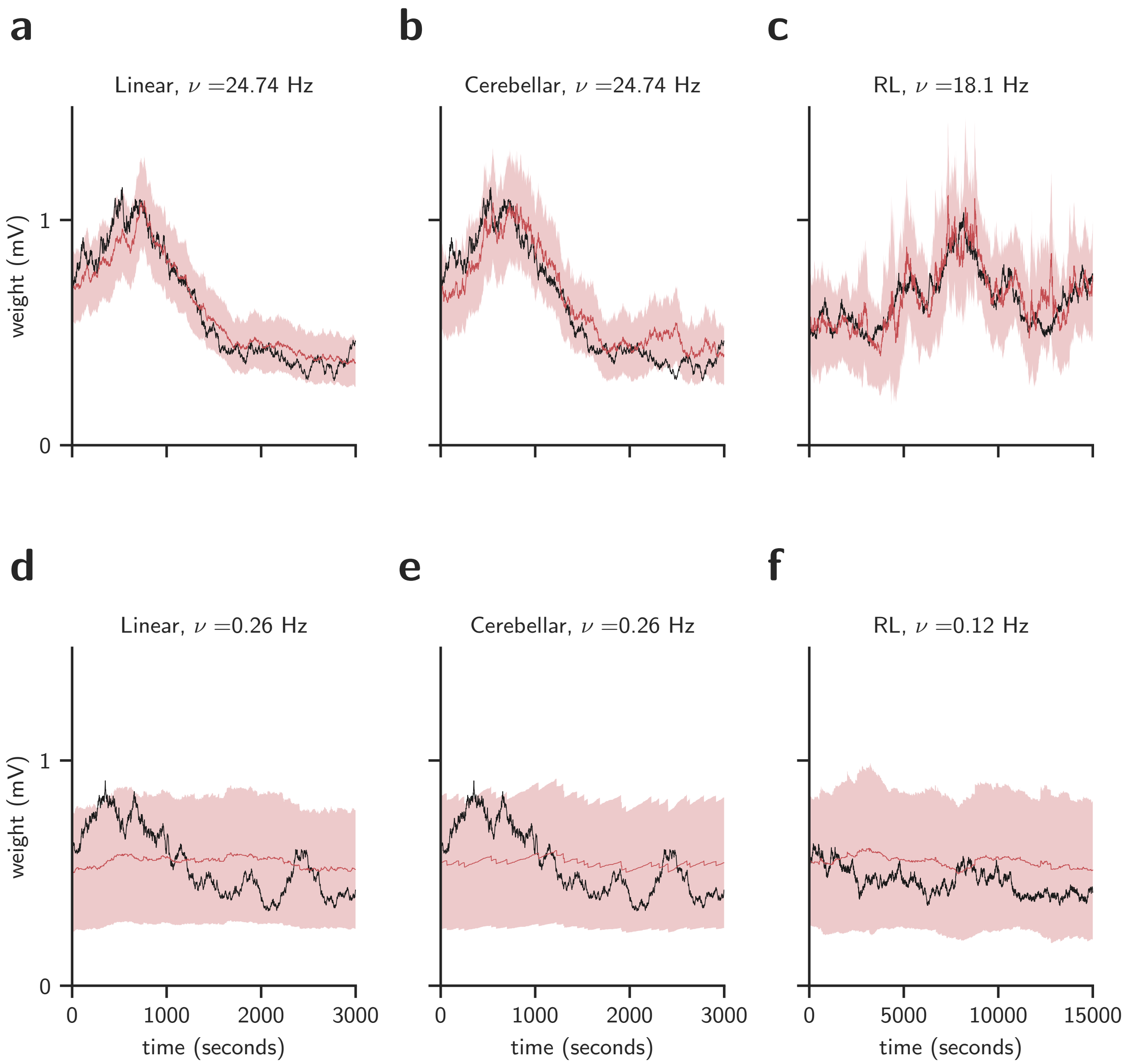

Figure 2: Bayesian learning rules track the target weight and estimate uncertainty. The black line is the target weight, the red line is the mean of the inferred distribution, and the red area represents 95% confidence intervals of the inferred distribution. Panels a-c correspond to the highest presynaptic firing rate used in the simulations; panels d-f to the lowest. Consistent with our analysis (see in particular Eq. 9), higher presynaptic firing rate resulted in lower uncertainty. **a** and **d**. Linear feedback, $f_{\text{lin}} = \delta + \xi_\delta$. **b** and **e**. Cerebellar learning, $f_{\text{cb}} = \Theta(\delta + \xi_\delta - \theta)$. **c** and **f**. Reinforcement learning, $f_{\text{rl}} = -|\delta + \xi_\delta|$. See Supplementary Math Note, Sec. S5, for simulation details. Note while the red lilnes are all plotted at the same thickness, the greater variability in the lower plots may make those lines appear thicker.

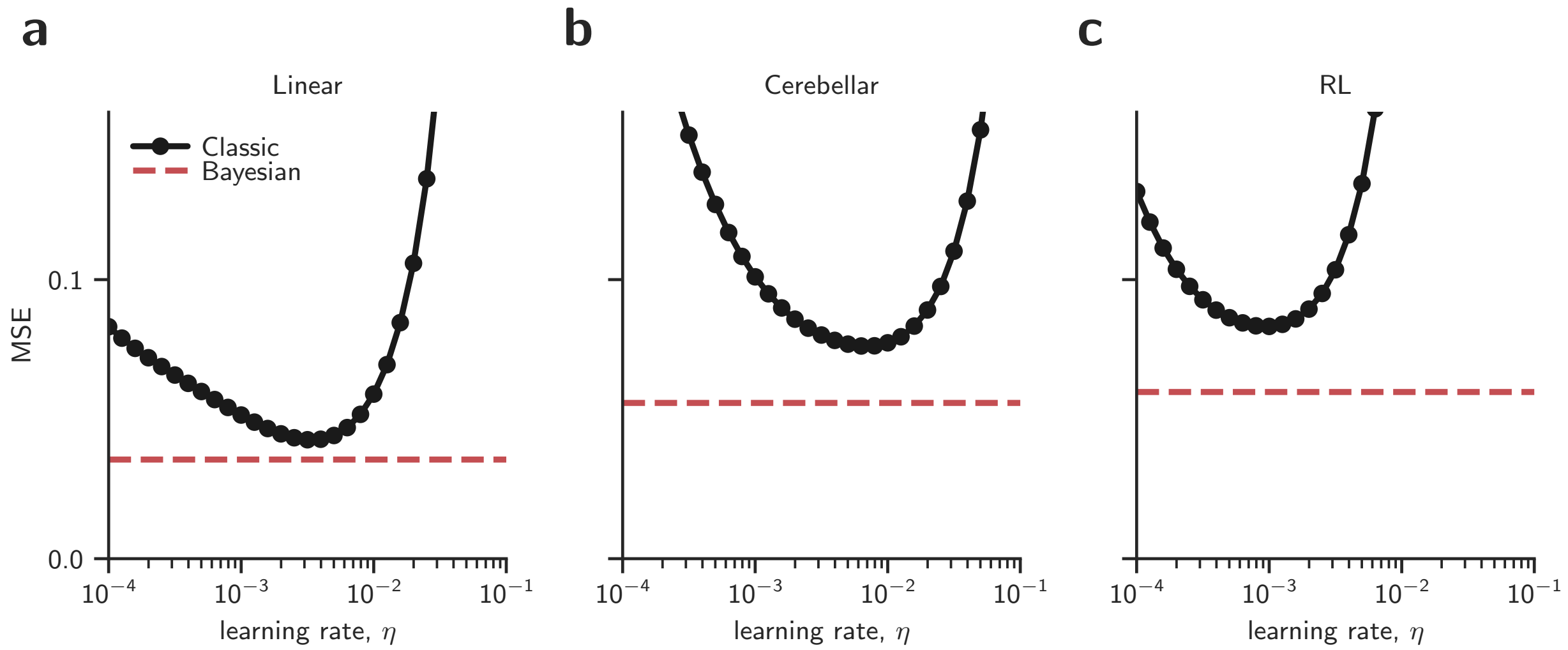

Figure 3: Bayesian learning rules exhibit lower error than classical ones. Red: mean squared error between the target and actual membrane potential for the Bayesian learning rules; black: mean squared error for the classical rules. **a**. Linear feedback, $f_{\text{lin}} = \delta + \xi_\delta$. **b**. Cerebellar learning, $f_{\text{cb}} = \Theta\left(\delta + \xi_\delta - \theta\right)$. **c**. Reinforcement learning, $f_{\text{rl}} = -|\delta + \xi_\delta|$. See Supplementary Math Note, Sec. S5, for simulation details.

We focused on this case primarily to illustrate our method in the simplest possible setting. In reality, however, the brain needs to optimize some cost function based on a feedback signal applied to a recurrent neural network. To investigate Bayesian plasticity in this, more realistic, regime, we trained the output weights of a recurrent neural network to produce a target function, using as a feedback signal the difference between the target function and its network estimate (Fig. 4a). The learning rules are very similar to Eq. (4) (see Methods, Sec. M2.3). However, the target weights are not known, so we cannot compare the inferred weights to the target weights, as we did in Fig. 2. We can, however, compare the mean squared error between the target and actual membrane potential, as in Fig. 3. Bayesian plasticity does indeed outperform classical learning rules (Figs. 4b and c). Moreover, the effect is much larger than in Fig. 3: the mean squared error is about an order of magnitude smaller for the Bayesian than for the classical learning rule (note the log scale in Fig. 4c), a result that was highly robust to model mismatch (Supplementary Math Note, Fig. S.12). These simulations suggest that taking into account weight uncertainty has a much larger effect in networks than in single neurons.

Figures 3 and 4 indicate that there is a clear advantage to using uncertainty to adjust learning rates. But does the brain do this? Addressing that question will require a new generation of plasticity experiments. At present, in typical plasticity experiments only changes in weights are measured; to test our hypothesis, it will be necessary to measure changes in learning rates, and at the same time determine how those changes are related to the synapse's uncertainty. This presents two challenges. First, measuring changes in learning rates is difficult, as weights must be monitored over long periods of time and under natural conditions, preferably *in vivo*. Second, we cannot measure the synapse's uncertainty directly. Here we discuss two approaches to overcoming these challenges.

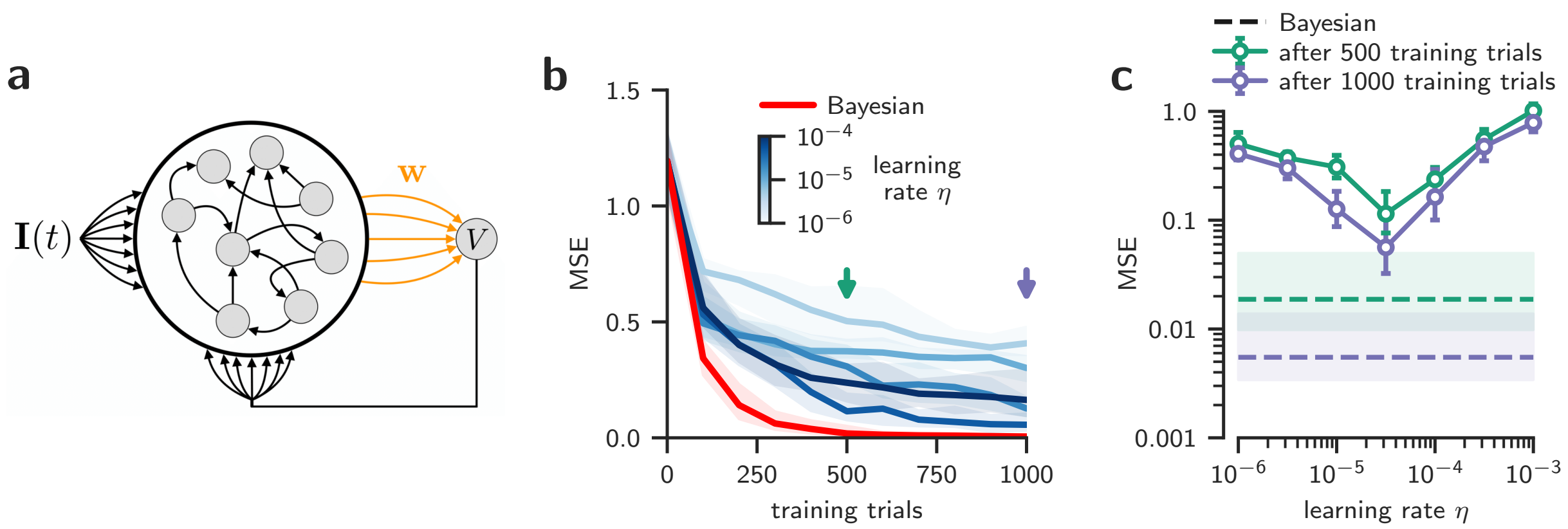

Figure 4: Recurrent neural network. **a**. Schematic of the circuit. $I(t)$ is the input (used to initialize activity) and $\mathbf{w}$ corresponds to the learned output weights. The feedback weights (black arrows from $V$ to the recurrent network) are fixed, as are the recurrent weights. During learning, the output of the network, $V(t)$, is compared to the target output, $V_{\text{tar}}(t)$, and the error is used to update the output weights, $\mathbf{w}$. At test time, the target output is not fed back to the circuit. **b**. Learning curves, measured using mean squared error, for Bayesian and classical learning rules (red and blue, respectively, at a range of learning rates for the classical rule). Although the initial improvement in performance for the Bayesian and classical learning rules was about the same, after 100 time steps Bayesian learning became much more efficient. The arrows correspond to the number of time steps used for the comparison in panel c. **c**. Mean squared error versus the learning rate of the classical rule. Solid lines: classical learning rules; dashed lines: Bayesian learning rules. The mean squared error for the Bayesian learning rule was about an order of magnitude smaller than for the classical one. In panels b and c we plot the median, taken over $n = 400$ network/target pairs; error bars are 95% confidence intervals, computed using the percentile bootstrap.

The first approach is indirect: use neural activity measured over long periods *in vivo* to estimate the uncertainty a synapse should have; then, armed with that estimate, test the prediction that the learning rate increases with uncertainty. To estimate the uncertainty a synapses should have, we take advantage of a general feature of essentially all learning rules: synapses get information only when the presynaptic neuron spikes. Consequently, the synapse's uncertainty should fall as the presynaptic firing rate increases. In fact, under mild assumptions, we can derive a very specific relationship: the relative change in weight under a plasticity protocol, $\Delta\mu_i/\mu_i$, should scale approximately as $1/\sqrt{\nu_i}$ where $\nu_i$ is the firing rate of the neuron presynaptic to synapse $i$,

$$\frac{\Delta\mu_i}{\mu_i} \propto \frac{1}{\sqrt{\nu_i}}\,, \tag{6}$$

a relationship that holds in our simulations for firing rates above about 1 Hz (see Supplementary Math Note, Fig. S.8, bottom row). In essence, firing rate is a proxy for uncertainty, with higher firing rate indicating lower uncertainty and vice versa. This prediction could be tested by observing neurons *in vivo*, estimating the presynaptic firing rates, then performing plasticity experiments to determine the relative change in synaptic strength, $\Delta\mu_i/\mu_i$.

The second approach is more direct, but it requires an additional hypothesis. While Bayesian Plasticity tells us how to compute the mean and variance of the weights, it does not tell us what weight to use when a spike arrives. But the Synaptic Sampling hypothesis does: it tells us that the mean and variance of the PSP amplitude should be equal to the mean and variance of the inferred distribution over the target weight,

$$\text{PSP mean} = \mu_i \tag{7a}$$

$$\text{PSP variance} = \sigma_i^2. \tag{7b}$$

Under our learning rules, the change in mean synaptic weight is proportional to the variance, $\sigma_i^2$ (Eq. 4a). Consequently, the relative change in weight $\Delta\mu_i/\mu_i$, is proportional to $\sigma_i^2/\mu_i$; combining this with Eq. (7) gives us

$$\frac{\Delta\mu_i}{\mu_i} \propto \frac{\text{PSP variance}}{\text{PSP mean}} \equiv \begin{matrix}\text{Normalized}\\ \text{Variability}\end{matrix}\,, \tag{8}$$

where we have defined the normalized variability to be the ratio of PSP variance to its mean. We verify that this relationship holds in simulations (see Supplementary Math Note, Fig. S.7).

Equation (8) implies that when the PSP variance is high, learning rates are also high. Testing that experimentally is technically difficult: it requires monitoring the PSP mean and variance for long periods *in vivo*, and comparing normalized variability to changes in the mean. However, such experiments are likely to be possible in the near future.

A more indirect approach based on this idea, for which we can apply current data, makes use of Eq. (6) to replace the left hand side of Eq. (8), $\Delta\mu_i/\mu_i$, with $1/\sqrt{\nu_i}$. This gives us

$$\begin{matrix}\text{Normalized}\\ \text{Variability}\end{matrix} \propto \frac{1}{\sqrt{\nu_i}}\,. \tag{9}$$

This is intuitively sensible: as discussed above, higher presynaptic firing rates means the synapse is more certain, and Synaptic Sampling states that higher certainty should reduce the observed variability. This relationship can be tested by estimating presynaptic firing rates *in vivo*, and comparing them to the normalized variability measured using paired recordings. Such data can be extracted from experiments by Ko and colleagues [15]. In those experiments, calcium signals in mouse visual cortex were recorded *in vivo* under a variety of stimulation conditions, which provided an estimate of the firing rate of each imaged neuron; subsequently, whole cell recordings of pairs of identified neurons were made *in vitro*, and the mean and variance of the PSPs were measured. In Fig. 5 we plot the normalized variability versus the firing rate on a log-log scale; on this scale, our theory predicts a slope of $-1/2$ (red line). The normalized variability does indeed decrease as the firing rate increases (blue line), ($p < 0.003$), and the slope is not significantly different from the predicted value of $-1/2$ ($p = 0.57$). This pattern is broadly matched by simulations, at least at sufficiently high firing rate (Supplementary Math Note, Fig. S.8, top row).

An alternative explanation for this result is that increases in firing rate reduce the normalized variability because of short term effects on release probability. The release probability, denoted $p_r$, scales the variance of the PSP by a factor of $p_r(1 - p_r)$ and the mean by a factor of $p_r$, so the normalized variability (the variance divided by the mean) scales as $1-p_r$. Consequently, an increase in release probability with firing rate would explain Fig. 5. Such increases do indeed occur [16]. However, much more common — especially in rodent layer 2/3, where these experiments were performed — is a decrease in release probability with firing rate [17, 18]. Thus, short term synaptic plasticity would typically lead to an increase, not a decrease, in the normalized variability when firing rate increases; the opposite of what we see experimentally.

# 3 Discussion

We proposed that synapses do not just keep track of point estimates of their weights, as they do in classical learning rules; they also keep track of their uncertainty. They then use that uncertainty to set learning rates: the higher the uncertainty, the higher the learning rate. This allows different synapses to have different learning rates, and leads to learning rules that allow synapses to exploit all locally available information. This in turn leads to better performance, as measured by mean squared error (Figs. 3 and 4b,c). It also leads to faster learning. That is implicit in Fig. 3 (because the target weights drift, fast learning is essential for achieving low mean squared error) and it is explicit in Fig. 4b (compare red to blue curves).

The critical difference between our learning rules and classical ones is that the learning rates themselves undergo plasticity. We derived three rules, based on three different assumptions about the feedback signal received by the neuron, and in all cases the updates for the mean had the flavor of a classical rule: the change in the mean weight was a function of the presynaptic activity and an error signal. Other assumptions about the feedback signal are clearly possible, and our method can generate a broad range of learning rules. Whether or not they can generate all rules that have been observed experimentally is an avenue for future research.

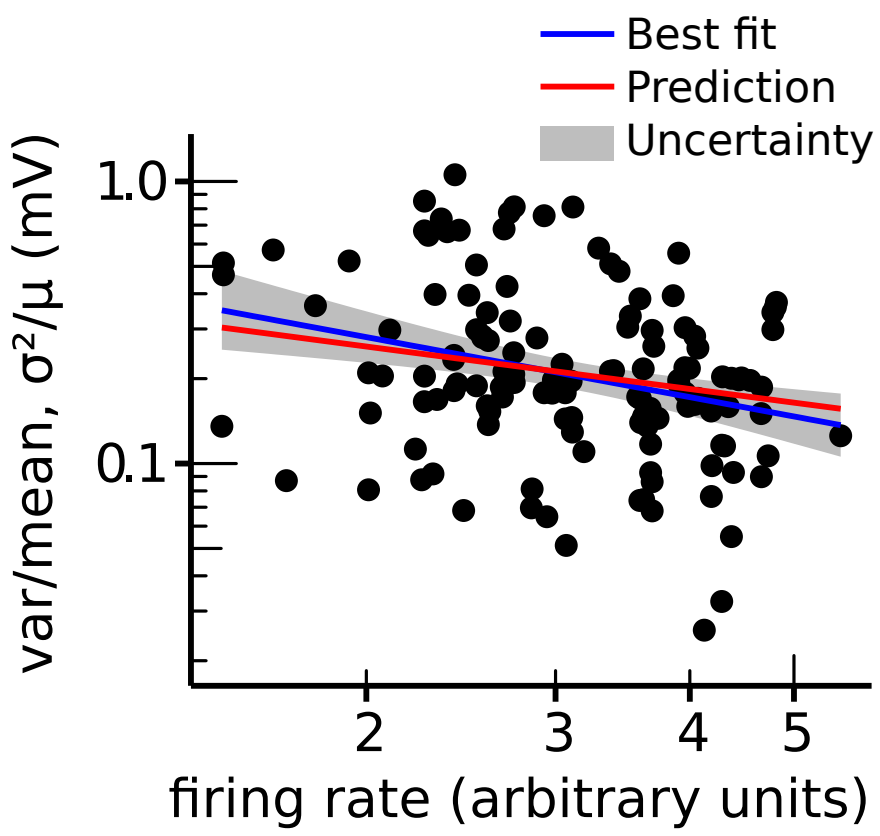

Figure 5: Normalized variability (the ratio of the PSP variance to the mean) versus presynaptic firing rate as a diagnostic of our theory; data supplied to us by the authors of [15] (see Supplementary Math Note, Sec. S6.4). The red line, which has a slope of $-1/2$, is our prediction (the intercept, for which we do not have a prediction, was chosen to give the best fit to the data). The blue line is fit by linear regression($n = 136$ points), and the gray region represents 2 standard errors. The slope of the blue line, -0.62, is statistically significantly different from 0 ($p < 0.003$, $t$-test) and not significantly different from $-1/2$ ($p = 0.57$, $t$-test; assumes normality which was not formally tested). The firing rate was measured by taking the average signal from a spike deconvolution algorithm [45]. Units are arbitrary because the scale factor relating the average signal from the deconvolution algorithm and the firing rate is not exactly one [46]. Data from layer 2/3 of mouse visual cortex [15].

The hypothesis that synapses keep track of uncertainty, which we refer to as the Bayesian Plasticity hypothesis, makes the general prediction that learning rates, not just synaptic strengths, are a function of pre and postsynaptic activity — something that should be testable with the next generation of plasticity experiments. In particular, it makes a specific prediction about learning rates *in vivo*: learning rates should vary across synapses, being higher for synapses with lower presynaptic firing rates.

We also made a second, independent, hypothesis, Synaptic Sampling. This hypothesis states that the variability in PSP size associated with a particular synapse matches the uncertainty in the strength of that synapse. This allows synapses to communicate their uncertainty to surrounding circuitry — information that is critical if the brain is to monitor the accuracy of its own computations. The same principle has been applied to neural activity, where it is known as the neural sampling hypothesis [19, 20, 21, 22], which posits that variability in neural activity matches uncertainty about the state of the external world. The neural sampling hypothesis meshes well with synaptic sampling: uncertainty in the weights increases uncertainty in the current estimate of the state of the world, and likewise, variability in the weights increase variability in neural activity (see Supplementary Math Note, Sec. S7). While there is some experimental evidence for the neural sampling hypothesis [21, 23, 22, 24, 25], it has not been firmly established. Whether other proposals for encoding probability distributions with neural activity, such as probabilistic population codes [3, 26], can be combined with Synaptic Sampling is an open question.

By combining our two hypotheses, we were able to make additional predictions. These focused on what we call the normalized variability — the ratio of the variance in PSP size to the mean. First, we predicted that plasticity should increase with normalized variability, which remains to be tested. Second, we predicted that normalized variability should decrease with presynaptic firing rate. Reanalyzing data from [14], we provided evidence that this is indeed the case.

In machine learning, the idea that it is advantageous to keep track of the distribution over weights has a long history [27, 28, 29]. Especially relevant is a recent study in which, as in our scheme, learning rates were reduced when certainty was high [30]. However, rather than updating the uncertainty on every time step, as we do, updating occurred only when there was a change in the task. This occurs on the timescale of minutes to hours; not the millisecond timescale on which uncertainty is updated in our model. Nevertheless, this approach worked well in settings in which deep networks had to learn multiple tasks.

In neuroscience, weight uncertainty was first explored in the context of reinforcement learning [31]. In that work, the weights related sensory stimuli to rewards, and weight correlations that developed due to Bayesian learning provided an exceptionally elegant explanation of backward blocking. The idea lay dormant for over a decade, until it was rediscovered with a slightly different focus, one in which knowledge of weight uncertainty is critical for knowledge of computational uncertainty [3]. Several theoretical studies followed. The first of those [32] bore some resemblance to ours, in that weights were sampled from a distribution. However, the timescale for sampling was hours rather than milliseconds; too slow to explain the spike-to-spike variability in PSP size that is ubiquitous in the brain. More recently, Hiratani and Fukai [33] postulated that the multiple synaptic contacts per connection observed in cortex provides a scaffolding for constructing a non-parametric estimate of the probability distribution over synaptic strength. Weight uncertainty has also been applied to drift diffusion models [34], using methods similar to those in [31]; the main difference was that the reward was binary (correct or incorrect) rather than continuous. Finally, recent work proposed that short-term plasticity is also governed by a Bayesian updating process [35]. It will be interesting to determine which combination of these schemes is used by the brain.

If the Bayesian Plasticity hypothesis is correct, synapses would have to keep track of, and store, two variables: the mean, as is standard, but also the variance (or, equivalently, the learning rate), which is not. The complexity of synapses [36, 37, 38], and their ability to use non-trivial learning rules (e.g., synaptic tagging, in which activity at a synapse "tags" it for future long term changes in strength [39, 40, 41], and metaplasticity, in which the learning rate can be modified by synaptic activity without changing the synaptic strength [42, 43, 44]), suggests that representing uncertainty — or learning rate — is quite possible. It will be nontrivial, but important, to work out how.

Our framework has several implications, both for the interpretation of neurophysiological data and for future work. First, under the Synaptic Sampling hypothesis, PSPs are necessarily noisy. Consequently, noise in synapses (e.g., synaptic failures) is a feature, not a bug. We thus provide a normative theory for one of the major mysteries in synaptic physiology: why neurotransmitter release is probabilistic. Second, our approach allows us to derive local, biologically plausible learning rules, no matter what information is available at the synapse, and no matter what the statistics of the synaptic input. Thus, our approach provides the flexibility necessary to connect theoretical approaches based on optimality to complex biological reality.

In neuroscience, Bayes theorem is typically used to analyze high level inference problems, such as decision-making under uncertainty. Here we demonstrated that Bayes' theorem, being the optimal way to solve any inference problem, big or small, could be implemented in perhaps the smallest computationally relevant element in the brain: the synapse.

# Acknowledgments

LA and PEL were supported by the Gatsby Charitable Foundation; PEL was also supported by the Wellcome Trust (110114/Z/15/Z); JJ and JPP were supported by the Swiss National Science Foundation (PP00P3 150637); JAM was supported by UCL Graduate Research and UCL Overseas Research Scholarships; AP was supported by grants from the Simons Collaboration for the Global Brain and the Swiss National Foundation (31003A 165831).

# Author Contributions

AP and PEL were involved in the initial formulation of the problem. LA and PEL were involved in the theoretical development. LA, JJ and JM did the simulations and data analysis. PEL, JP and AP were involved in writing the manuscript.

# Competing interests

The authors declare no competing interests.

# References

[1] T. Poggio. "A theory of how the brain might work". In: *Cold Spring Harbor Symposia on Quantitative Biology*. Vol. 55. Cold Spring Harbor Laboratory Press, 1990.

[2] D. C. Knill and W. Richards. *Perception as Bayesian Inference*. Cambridge University Press, 1996.

[3] A. Pouget et al. "Probabilistic brains: knowns and unknowns". In: *Nature Neuroscience* 16.9 (2013), pp. 1170–1178.

[4] L. Aitchison. "Bayesian filtering unifies adaptive and non-adaptive neural network optimization methods". In: *NeurIPS* (2020).

[5] S. J. Tripathy et al. "Brain-wide analysis of electrophysiological diversity yields novel categorization of mammalian neuron types". In: *Journal of Neurophysiology* 113.10 (2015), pp. 3474–3489.

[6] M. Schiess, R. Urbanczik, and W. Senn. "Somato-dendritic Synaptic Plasticity and Error-backpropagation in Active Dendrites". In: *PLoS Computational Biology* 12.2 (2016), e1004638.

[7] J. Bono and C. Clopath. "Modeling somatic and dendritic spike mediated plasticity at the single neuron and network level". In: *Nature Communications* 8 (2017), p. 706.

[8] J. Sacramento et al. "Dendritic cortical microcircuits approximate the backpropagation algorithm". In: *Advances in Neural Information Processing Systems 31*. Ed. by S. Bengio et al. Curran Associates, Inc., 2018, pp. 8721–8732.

[9] B. Illing, W. Gerstner, and J. Brea. "Biologically plausible deep learning – But how far can we go with shallow networks?" In: *Neural Networks* 118 (2019), pp. 90–101.

[10] M. Akrout et al. "Deep Learning without Weight Transport". In: *arXiv:1904.05391* (2019).

[11] M. Ito, M. Sakurai, and P. Tongroach. "Climbing fibre induced depression of both mossy fibre responsiveness and glutamate sensitivity of cerebellar Purkinje cells". In: *Journal of Physiology* 324.1 (1982), pp. 113–134.

[12] J. Eccles, R Llinas, and K Sasaki. "The excitatory synaptic action of climbing fibres on the Purkinje cells of the cerebellum". In: *Journal of Physiology* 182.2 (1966), pp. 268–296.

[13] B. Widrow and M. E. Hoff. “Adaptive switching circuits”. In: *Office of Naval Research Technical Report 1553-1* (1960).

[14] P. Dayan and L. F. Abbott. *Theoretical Neuroscience*. Cambridge, MA: MIT Press, 2001.

[15] H. Ko et al. “The emergence of functional microcircuits in visual cortex”. In: *Nature* 496.7443 (2013), pp. 96–100.

[16] A. M. Thomson. “Presynaptic frequency- and pattern-dependent filtering”. In: *Journal of Computational Neuroscience* 15.2 (2003), pp. 159–202.

[17] M. V. Tsodyks and H. Markram. “The neural code between neocortical pyramidal neurons depends on neurotransmitter release probability”. In: *Proceedings of the National Academy of Sciences* 94.2 (1997), pp. 719–723.

[18] A. Maffei and G. G. Turrigiano. “Multiple modes of network homeostasis in visual cortical layer 2/3”. In: *Journal of Neuroscience* 28.17 (2008), pp. 4377–4384.

[45] J. T. Vogelstein et al. “Fast nonnegative deconvolution for spike train inference from population calcium imaging”. In: *Journal of Neurophysiology* 104.6 (2010), pp. 3691–3704.

[46] A. M. Packer et al. “Simultaneous all-optical manipulation and recording of neural circuit activity with cellular resolution in vivo”. In: *Nature Methods* 12.2 (2015), pp. 140–146.

[19] P. O. Hoyer and A. Hyvarinen. “Interpreting neural response variability as Monte Carlo sampling of the posterior”. In: *Advances in Neural Information Processing Systems*. 2003, pp. 293–300.

[20] J. Fiser et al. “Statistically optimal perception and learning: from behavior to neural representations”. In: *Trends in Cognitive Sciences* 14.3 (2010), pp. 119–130.

[21] P. Berkes et al. “Spontaneous Cortical Activity Reveals Hallmarks of an Optimal Internal Model of the Environment”. In: *Science* 331.6013 (2011), pp. 83–87.

[22] G. Orbán et al. “Neural Variability and Sampling-Based Probabilistic Representations in the Visual Cortex”. In: *Neuron* 92.2 (2016), pp. 530–543.

[23] R. M. Haefner, P. Berkes, and J. Fiser. “Perceptual decision-making as probabilistic inference by neural sampling”. In: *Neuron* 90.3 (2016), pp. 649–660.

[24] L. Aitchison and M. Lengyel. “The Hamiltonian Brain: Efficient Probabilistic Inference with Excitatory-Inhibitory Neural Circuit Dynamics”. In: *PLoS Computational Biology* 12.12 (), e1005186.

[25] R. D. Lange and R. M. Haefner. "Task-induced neural covariability as a signature of approximate Bayesian learning and inference". In: *bioRxiv* (2020).

[26] W. J. Ma et al. "Bayesian inference with probabilistic population codes". In: *Nature Neuroscience* 9.11 (2006), pp. 1432–1438.

[27] W. L. Buntine and A. S. Weigend. "Bayesian backpropagation". In: *Complex systems* 5.6 (1991), pp. 603–643.

[28] D. J. MacKay. "A practical Bayesian framework for backpropagation networks". In: *Neural Computation* 4.3 (1992), pp. 448–472.

[29] C. Blundell et al. "Weight uncertainty in neural networks". In: *arXiv:1505.05424* (2015).

[30] J. Kirkpatrick et al. "Overcoming catastrophic forgetting in neural networks". In: *Proceedings of the National Academy of Sciences* 106.25 (2016), pp. 10296–10301.

[31] P. Dayan and S. Kakade. "Explaining Away in Weight Space". In: *Advances in Neural Information Processing Systems 13*.

[32] D. Kappel et al. "Network Plasticity as Bayesian Inference". In: *PLoS computational biology* 11.11 (2015), e1004485.

[33] N. Hiratani and T. Fukai. "Redundancy in synaptic connections enables neurons to learn optimally". In: *Proceedings of the National Academy of Sciences* 115.29 (2018), E6871–E6879.

[34] J. Drugowitsch et al. "Learning optimal decisions with confidence". In: *bioRxiv* (2019).

[35] J.-P. Pfister, P. Dayan, and M. Lengyel. "Synapses with short-term plasticity are optimal estimators of presynaptic membrane potentials". In: *Nature Neuroscience* 13.10 (2010), pp. 1271–1275.

[36] H. Kasai, N. Takahashi, and H. Tokumaru. "Distinct initial SNARE configurations underlying the diversity of exocytosis". In: *Physiological Reviews* 92 (2012), pp. 1915–1964.

[37] T. C. Südhof. "The presynaptic active zone". In: *Neuron* 75 (2012), pp. 11–25.

[38] K. Michel et al. "The presynaptic active zone: A dynamic scaffold that regulates synaptic efficacy". In: *Experimental Cell Research* 335 (2015), pp. 157–164.

[39] U. Frey and R. G. Morris. "Synaptic tagging and long-term potentiation". In: *Nature* 385.6616 (1997), pp. 533–536.

[40] R. L. Redondo and R. G. M. Morris. "Making memories last: the synaptic tagging and capture hypothesis". In: *Nature Reviews Neuroscience* 12.1 (2011), pp. 17–30.

[41] T. Rogerson et al. "Synaptic tagging during memory allocation". In: *Nature Reviews Neuroscience* 15.3 (2014), pp. 157–169.

[42] W. C. Abraham and M. F. Bear. "Metaplasticity: the plasticity of synaptic plasticity". In: *Trends in Neurosciences* 4 (1996), pp. 126–130.

[43] W. C. Abraham. "Metaplasticity: tuning synapses and networks for plasticity". In: *Nature Reviews Neuroscience* 9.5 (2008), p. 387.

[44] S. R. Hulme et al. "Mechanisms of heterosynaptic metaplasticity". In: *Philosophical Transactions of the Royal Society of London. Series B, Biological Sciences* 369.1633 (2014), p. 20130148.

# Methods

Here we provide a complete description of our model (Sec. M1) and sketch the derivation of the learning rules (Sec. M2).

## M1 Description of our model

In the main text we specified how the membrane potential depends on the weights and incoming spikes (Eq. 1) and how the target membrane potential depends on the target weights and incoming spikes (Eq. 5), and we defined the error signal (Eq. 2). In this section we describe how target weights, $w_{\text{tar},i}$, the weights, $w_i$, the and the spikes, $x_i$, are generated.

### M1.1 Target weights

The target weights are the weights that in some sense optimize the performance of the animal. We do not expect these weights to remain constant over time, for two reasons. First, both the state of the world and the organism change over time, thus changing the target weights. Second, we take a local, single neuron view to learning, and define the target weights on a particular neuron to be the optimal weights given the weights on all the other neurons in the network. Consequently, as the weights of surrounding neurons change due to learning, the target weights on our neuron also change. While these changes may be quite systematic, to a single synapse deep in the brain they are likely to appear random.

Motivated by this last observation, in our model we assume that the target weights evolve according to a random process. To ensure that the weights do not change sign, we work in log space, and on each time step we add a small amount of noise to the log of the target weights. And to ensure that the weights do not become too small or too large, we add a small drift toward a prior log weight. Specifically, defining

$$\lambda_{\text{tar},i} = \log |w_{\text{tar},i}| \tag{M.10}$$

(note the absolute value sign, which allows the weights to be either positive or negative), we let $\lambda_{\text{tar},i}$ evolve according to

$$\lambda_{\text{tar},i}(t+1) = \lambda_{\text{tar},i}(t) - \frac{\lambda_{\text{tar},i}(t) - m_{\text{prior}}}{\tau} + \sqrt{\frac{2s_{\text{prior}}^2}{\tau}}\,\xi_{\text{tar},i} \tag{M.11}$$

where $m_{\text{prior}}$ and $s_{\text{prior}}^2$ are the prior mean and variance of $\lambda_{\text{tar},i}(t)$, $\tau$ (which is dimensionless) is the characteristic number of steps over which $\lambda_{\text{tar},i}(t)$ changes, and $\xi_{\text{tar},i}$ is a zero mean, unit variance Gaussian random variable.

We chose the noise process described in Eq. (M.11) for three reasons. First, $w_{\text{tar},i}$ is equal to either $+e^{\lambda_{\text{tar},i}}$ (for excitatory weights) or $-e^{\lambda_{\text{tar},i}}$ (for inhibitory weights), and thus cannot change sign as $\lambda_{\text{tar},i}$ changes with learning. Consequently, excitatory weights cannot become inhibitory, and *vice versa*, so Dale's law is preserved. Second, spine sizes obey this stochastic process [1], and while synaptic weights are not spine sizes, they are correlated [2]. Third, this noise process gives a log-normal stationary distribution of weights, as is observed experimentally [3].

The parameters that determine how the weights drift, $m_{\text{prior}}$ and $s_{\text{prior}}^2$, were set to the mean and variance of measured log-weights using data from Ref. [3] (Supplementary Math Note, Sec. S6.1). We used a time step of 10 ms, within the range of measured membrane time-constants. For the linear and cerebellar models we set $\tau$ to $10^5$; for reinforcement learning we set $\tau$ to $5 \times 10^5$. These values were chosen so that uncertainty roughly matched observed variability (Supplementary Math Note, Sec. S6.3). For the recurrent network we do not know the target weights, so we do not know the drift rate. Nor do we know the effective drift associated with the fact that the optimal weight on one synapses changes as the surrounding circuit changes. We therefore tried different drifts in our simulations (data not shown). We found that near zero drift was optimal, so we set $\tau$ to $\infty$.

### M1.2 Synaptic weights

Our inference algorithm computes a distribution over the target weights. Given that distribution, though, there is nothing in the Bayesian Plasticity hypothesis that tells us how to set the weights when a spike arrives. That is where the Sampling Hypothesis comes in: it tells us to sample the weights, $w_i$, from the posterior,

$$w_i = e^{m_i + s_i \xi_i}, \tag{M.12}$$

where $m_i$ and $s_i$ are the mean and standard deviation of the posterior distribution over the log weights, and $\xi_i$ is a zero mean, unit variance Gaussian random variable. The mean and variance of $w_i$ under Eq. (M.12), for which we use $\mu_i$ and $\sigma_i^2$, respectively, are the standard expressions for the mean and variance of a log-normal distribution,

$$\mathbb{E}\left[w_i|\mathcal{D}_i\right] \equiv \mu_i = e^{m_i + s_i^2/2} \tag{M.13a}$$

$$\mathrm{Var}\left[w_i|\mathcal{D}_i\right] \equiv \sigma_i^2 = \mu_i^2\left[e^{s_i^2} - 1\right] \tag{M.13b}$$

where $\mathcal{D}_i$ is the data seen by the synapse so far (see Eq. M.19 below).

In the main text, we compare our Bayesian learning rules to classical ones (see in particular Figs. 3 and 4). For classical rules there is no posterior to sample from, so we can not use Eq. (M.12). Consequently, for the the classical implementation of linear and cerebellar rules we do not sample, and for Bayesian learning we use $w_i = \mu_i$. The reinforcement learning rule, however, requires sampling, for both Bayesian and classical learning (see Eqs. M.47 and M.48). We thus assumed that the variance is proportional to the mean (as is the case for Poisson statistics). To find the constant of proportionality, denoted $k$, we use data from Ref. [3]; see Supplementary Math Note, Sec. S6.2 for details. A least squares fit to that data gives $k = 0.0877$. A naive way to implement this is to sample weights using $w_i = \mu_i + \sqrt{k\mu_i}\,\xi_i$ with $\xi_i \sim \mathcal{N}(0, 1)$. However, that allows $w_i$ to change sign, so instead we sample the weights using

$$w_i = \mu_i e^{\beta_i + \gamma_i \xi_i}\,, \tag{M.14}$$

and choose $\beta_i$ and $\gamma_i$ so that the mean and variance of $w_i$ are $\mu_i$ and $k\mu_i$, respectively. As is straightforward to show, these conditions are satisfied when $\beta_i$ and $\gamma_i$ are given by

$$\beta_i = -\frac{\log(1 + k/\mu_i)}{2} \tag{M.15a}$$

$$\gamma_i = \sqrt{\log(1 + k/\mu_i)}\,. \tag{M.15b}$$

### M1.3 Synaptic input

For linear, cerebellar and reinforcement learning, neurons receive input from $n$ presynaptic neurons, all firing at different rates. The firing rates, $\nu_i$ ($i$ labels presynaptic neuron), are drawn from a log-normal distribution, using a distribution that is intermediate between the narrow range found by some [4] and the broad range found by others [5]: a log-normal with median at 1 Hz and with 95% of firing rates being between 0.1 Hz and 10 Hz,

$$\log \nu_i \sim \mathcal{N}\left(0, \left(\log\sqrt{10}\right)^2\right) \tag{M.16}$$

with $\nu_i$ measured in Hz. On each time step, $x_i$ is drawn from a Bernoulli distribution (so it is either 0 or 1),

$$P(x_i) = (\nu_i \Delta t)^{x_i} (1 - \nu_i \Delta t)^{1-x_i}. \tag{M.17}$$

## M2 Learning rules

Here we outline how a synapse can infer a probability distribution over its target weights. This is done using a well-understood class, hidden Markov models, for which we can use a standard, two-step procedure: in the first step the synapse incorporates new data using Bayes theorem; in the second step it take into account random changes in the target weight.

While straightforward in principle, in practice there are two difficulties with this approach. The first is that it results in a joint distribution over all synaptic weights. It is unlikely, however, that synapses could store such a distribution: even with a Gaussian approximation, for $n$ synapses there are about $n^2/2$ parameters. And it is even more unlikely that they could compute it, as that would require communication among synapses on different dendritic branches. We thus assume that each synapse performs probabilistic inference based only on the data available to it. This makes each synapse locally optimal, and it allows us to derive local learning rules. It is potentially the most important theoretical advance of our analysis. And within the Bayesian framework it is straightforward: each synapse simply integrates over the uncertainty in the target weights of all the other synapses. Nonetheless, this is an unusual approach, and further work is necessary to understand its theoretical properties.

The second difficulty is that even with the local approximation, inference is intractable, as it requires pointwise multiplication of probability distributions and a convolution (see Eqs. M.20 and M.21 below). To remedy this, we approximate the true distribution by a simpler one, a log-normal. The log-normal distribution was chosen for two reasons: it prevent synapses from changing sign, so Dale's law is respected; and it matches the distribution of the target weights, Eq. (M.11), so it produces the correct distribution in the absence of presynaptic spikes.

### M2.1 Single neuron learning rules: general formalism

The goal of a synapse is to compute the probability distribution over synaptic strength given data up to the last time step. Here the data – assumed local, as just discussed – consists of the feedback signal, $f$ (shorthand for $f_{\text{lin}}$, $f_{\text{cb}}$ or $f_{\text{rl}}$), the presynaptic input, $x_i$, and the actual weight, $w_i$. To reduce clutter, we use $d_i(t)$ to denote the data at time $t$,

$$d_i(t) \equiv \{f(t), x_i(t), w_i(t)\}, \tag{M.18}$$

and $\mathcal{D}_i(t)$ to denote past data,

$$\mathcal{D}_i(t) \equiv \{d_i(t), d_i(t-1), d_i(t-2), \dots\}. \tag{M.19}$$

With this notation, the goal of the synapse is to compute $P(\lambda_{\text{tar},i}(t+1)|\mathcal{D}_i(t))$ in terms of $P(\lambda_{\text{tar},i}(t)|\mathcal{D}_i(t-1))$. To reduce clutter even further, here and in what follows all quantities without an explicitly specified time index are evaluated at time step $t$; thus, we will derive an update rule for $P(\lambda_{\text{tar},i}(t+1)|\mathcal{D}_i)$ in terms of $P(\lambda_{\text{tar},i}|\mathcal{D}_i(t-1))$.

Making, as discussed above, the approximation that synapses perform inference based only on local information, the first step in the derivation of the update rule, incorporating new data using Bayes theorem, gives us

$$P(\lambda_{\text{tar},i}|\mathcal{D}_i) = P(\lambda_{\text{tar},i}|d_i, \mathcal{D}_i(t-1)) \propto P(d_i|\lambda_{\text{tar},i})\, P(\lambda_{\text{tar},i}|\mathcal{D}_i(t-1)) \tag{M.20}$$

where we used the Markov property: $P(d_i|\lambda_{\text{tar},i}, \mathcal{D}_i(t-1)) = P(d_i|\lambda_{\text{tar},i})$. (Recall that $\lambda_{\text{tar},i}$ is the log of the absolute value of the $i^{\text{th}}$ target weight, $w_{\text{tar},i}$; see Eq. M.10.) In the second step, the synapse takes into account random changes in the target weight,

$$P(\lambda_{\text{tar},i}(t+1)|\mathcal{D}_i) = \int d\lambda_{\text{tar},i} P(\lambda_{\text{tar},i}(t+1)|\lambda_{\text{tar},i})\, P(\lambda_{\text{tar},i}|\mathcal{D}_i)\,. \tag{M.21}$$

The conditional distribution, $P(\lambda_{\text{tar},i}(t+1)|\lambda_{\text{tar},i})$, can be extracted from Eq. (M.11). Combining both steps takes us from the distribution at time $t$, $P(\lambda_{\text{tar},i}|\mathcal{D}_i(t-1))$, to the distribution at the time $t+1$, $P(\lambda_{\text{tar},i}(t+1)|\mathcal{D}_i)$.

To make progress analytically, we approximate the true distribution by a log-normal one with mean $m_i$ and variance $s_i^2$; that is, we assume that

$$\lambda_{\text{tar},i}|\mathcal{D}_i(t-1) \sim \mathcal{N}\left(m_i, s_i^2\right). \tag{M.22}$$

This is the quantity the synapse needs when it sets the actual weight, $w_i$. (Recall that quantities with no explicit time dependence are to be evaluated at time $t$; thus, the left hand side is the probability distribution over $\lambda_{\text{tar},i}(t)$ given data up to the previous time step.)

Finalizing the calculation requires two steps: 1) insert Eq. (M.22) into (M.20) and compute $P(\lambda_{\text{tar},i}|\mathcal{D}_i)$; 2) insert that into Eq. (M.21) and compute $P(\lambda_{\text{tar},i}(t+1)|\mathcal{D}_i)$. However, Eq. (M.20) takes us out of our log-normal model class. To remedy this we use Assumed Density Filtering [6], for which posteriors are taken to be log-normal with mean and variance chosen to produce the distribution closest to the true one, where "close" is measured by the KL-divergence between the true and log-normal distributions. This can be achieved by matching moments: the mean and variance of the "closest" log-normal distribution are

$$m_i = \mathbb{E}\left[\lambda_{\text{tar},i}|\mathcal{D}_i(t-1)\right] \tag{M.23a}$$

$$s_i^2 = \text{Var}\left[\lambda_{\text{tar},i}|\mathcal{D}_i(t-1)\right]\,. \tag{M.23b}$$

We will apply this first to Eq. (M.20). Taking the log of both sides of that equation gives

$$\log P(\lambda_{\text{tar},i}|\mathcal{D}_i) = L(\lambda_{\text{tar},i}) + \log P(\lambda_{\text{tar},i}|\mathcal{D}_i(t-1)) + \text{const} \tag{M.24}$$

where

$$L(\lambda_{\text{tar},i}) \equiv \log P(d_i|\lambda_{\text{tar},i}) \tag{M.25}$$

is the log likelihood of the data at time $t$ given the target weight; we suppress the dependence on $d_i$ to avoid clutter. Under our log-normal assumption, the second term on the right hand side of Eq. (M.24) is Gaussian in $\lambda_{\text{tar},i}$. Motivated by the fact that new data does not provide much information, we assume that the likelihood is a slowly varying function of the target weights. This allows us to make a Laplace approximation: we Taylor expand the log likelihood around $m_i$, the mean of $P(\lambda_{\text{tar},i}|\mathcal{D}_i(t-1))$, and work only to second order in $\lambda_{\text{tar},i} - m_i$. Also using Eq. (M.22), we have

$$\log P(\lambda_{\text{tar},i}|\mathcal{D}_i) = L'(m_i)(\lambda_{\text{tar},i} - m_i) + L''(m_i)\,\frac{(\lambda_{\text{tar},i} - m_i)^2}{2} - \frac{(\lambda_{\text{tar},i} - m_i)^2}{2s_i^2} + \text{const.} \tag{M.26}$$

The right hand side is now quadratic in $\lambda_{\text{tar},i}$. Consequently, $P(\lambda_{\text{tar},i}|\mathcal{D}_i)$ is Gaussian, with mean and variance given by

$$\mathbb{E}\left[\lambda_{\text{tar},i}|\mathcal{D}_i\right] = m_i + \frac{s_i^2 L'(m_i)}{1 - s_i^2 L''(m_i)} \approx m_i + s_i^2 L'(m_i) \tag{M.27a}$$

$$\text{Var}\left[\lambda_{\text{tar},i}|\mathcal{D}_i\right] = s_i^2 + \frac{s_i^4 L''(m_i)}{1 - s_i^2 L''(m_i)} \approx s_i^2 + s_i^4 L''(m_i)\,. \tag{M.27b}$$

To derive the approximation expressions, we assumed $s_i^2|L''(m_i)| \ll 1$. This holds in the limit of slowly varying log likelihood, which we assume throughout our analysis.

Equation (M.27) tells us how to incorporate new data; we now need to incorporate random drift, via the integral in Eq. (M.21). From Eq. (M.11), we see that $P(\lambda_{\text{tar},i}(t+1)|\lambda_{\text{tar},i})$ is Gaussian, so the integral is straightforward, and we have

$$m_i(t+1) = \left(1 - \frac{1}{\tau}\right)\mathbb{E}\left[\lambda_{\text{tar},i}|\mathcal{D}_i\right] + \frac{m_{\text{prior}}}{\tau} \tag{M.28a}$$

$$s_i^2(t+1) = \left(1 - \frac{1}{\tau}\right)^2 \text{Var}\left[\lambda_{\text{tar},i}|\mathcal{D}_i\right] + \frac{2s_{\text{prior}}^2}{\tau}\,. \tag{M.28b}$$

Inserting Eq. (M.27) into (M.28), and working to lowest nonvanishing order in $1/\tau$, $s_i L'(m_i)$ and $s_i^2 L''(m_i)$, we arrive at our final update equations,

$$\Delta m_i = s_i^2 L'(m_i) - \frac{m_i - m_{\text{prior}}}{\tau} \tag{M.29a}$$

$$\Delta s_i^2 = s_i^4 L''(m_i) - \frac{2(s_i^2 - s_{\text{prior}}^2)}{\tau} \tag{M.29b}$$

where $\Delta m_i \equiv m_i(t+1) - m_i$ and $\Delta s_i^2 \equiv s_i^2(t+1) - s_i^2$. Thus, to update the mean and variance, all we have to do is compute the log likelihood and take the first and second derivatives. In Sec. M2.2 below we sketch how to do that; additional details are provided in Supplementary Math Note, Sec. S3. Note that equality in these expressions (and many that follow) is shorthand for equality under the assumptions and approximations of our model.

### M2.2 Single neuron learning rules for our three models

According to the above analysis (see in particular Eq. M.29), to determine the update rules we just need the log likelihood of the current data, $d_i(t)$, given the error signal (either $f_{\text{lin}}$, $f_{\text{cb}}$ or $f_{\text{rl}}$). Computing it is nontrivial, as several approximations are required. However, it is not hard to get an intuitive understanding of its form.

Using Eq. (M.18) for the data, $d_i$, the likelihood – the probability of the data given $w_{\text{tar},i}$ – may be written

$$P(d_i|\lambda_{\text{tar},i}) = P(f|x_i, w_i, \lambda_{\text{tar},i})\, P(x_i, w_i|\lambda_{\text{tar},i}) \propto P(f|x_i, w_i, \lambda_{\text{tar},i}) \ , \tag{M.30}$$

where we are able to drop the term $P(x_i, w_i|\lambda_{\text{tar},i})$ because without an error signal, $x_i$ and $w_i$ do not provide any information about $\lambda_{\text{tar},i}$. For all of our feedback signals, $f$ is a function of $f_{\text{lin}}$ (see main text); we take advantage of this to write

$$P(f|x_i, w_i, \lambda_{\text{tar},i}) = \int df_{\text{lin}}\, P(f|f_{\text{lin}})\, P(f_{\text{lin}}|x_i, w_i, \lambda_{\text{tar},i}) \ . \tag{M.31}$$

We focus here on computing $P(f_{\text{lin}}|x_i, w_i, \lambda_{\text{tar},i})$, and leave the integral for Supplementary Math Note, Sec. S3.1. Using Eqs. (1), (2) and (5) from the main text, we have

$$f_{\text{lin}} = (w_{\text{tar},i} - w_i)x_i + \sum_{j\neq i}(w_{\text{tar},j} - w_j)x_j + \xi_V + \xi_\delta \, . \tag{M.32}$$

For synapse $i$, all the terms in the sum over $j$ are unobserved, and so correspond to noise. By the Central Limit Theorem (and the assumed independence of the synapses), that noise is Gaussian; we take the added noise, $\xi_V$ and $\xi_\delta$, to be Gaussian as well, with total variance

$$\sigma_0^2 \equiv \text{Var}\left[\xi_\delta\right] + \text{Var}\left[\xi_V\right] \ . \tag{M.33}$$

Consequently, we may write

$$f_{\text{lin}}|w_i, x_i, \lambda_{\text{tar},i} \sim \mathcal{N}\left((w_{\text{tar},i} - w_i)x_i, \sigma_{\delta,i}^2\right) \tag{M.34}$$

where

$$\sigma_{\delta,i}^2 \equiv \text{Var}\left[\sum_{j\neq i}(w_{\text{tar},j} - w_j)\, x_j \,\middle|\, \mathcal{D}_i(t-1)\right] + \sigma_0^2 \, . \tag{M.35}$$

The quantity $\sigma^2_{\delta,i}$ depends on synapse, $i$. However, in the limit that there are a large number of synapses, that dependence is weak. We thus approximate it by including all terms in the sum over $j$, which we denote $\sigma^2_\delta$,

$$\sigma^2_{\delta,i} \approx \sigma^2_\delta \equiv \mathrm{Var}\left[\sum_j (w_{\mathrm{tar},i} - w_j)\, x_j \,\middle|\, \mathcal{D}_i(t-1)\right] + \sigma^2_0\,. \tag{M.36}$$

Under this approximation,

$$f_{\mathrm{lin}}|w_i, x_i, \lambda_{\mathrm{tar},i} \sim \mathcal{N}\left((\pm e^{\lambda_{\mathrm{tar},i}} - w_i)x_i, \sigma^2_\delta\right)\,. \tag{M.37}$$

For much of our analysis we use the value of $\sigma^2_\delta$ under the prior. That quantity, denoted $\sigma^2_{\delta 0}$, is given by

$$\sigma^2_{\delta 0} \equiv (\sigma^2_{\mathrm{prior}} + \sigma^2_{w,\,\mathrm{prior}}) \sum_j \nu_j \Delta t (1 - \nu_j \Delta t) + \sigma^2_0 \tag{M.38}$$

where the term $\nu_j \Delta t(1 - \nu_j \Delta t)$ comes from the Bernoulli statistics of $x_j$ (see Eq. M.17), and $\sigma^2_{\mathrm{prior}}$ and $\sigma^2_{w,\mathrm{prior}}$ are the variances of $w_{\mathrm{tar},i}$ and $w_i$ under the prior. The latter, $\sigma^2_{w,\mathrm{prior}}$, depends on whether or not we are sampling,

$$\sigma^2_{w,\,\mathrm{prior}} \equiv \begin{cases} \sigma^2_{\mathrm{prior}} & \text{Synaptic Sampling} \\ k\,\mu_{\mathrm{prior}} & \text{variance proportional to the mean (Sec. M1.2).} \end{cases} \tag{M.39}$$

The prior mean and variance of the weights in terms of the log weights (the quantities we have access to; see Table 1 below) are given by Eq. (M.13),

$$\mu_{\mathrm{prior}} \equiv e^{m_{\mathrm{prior}} + s^2_{\mathrm{prior}}/2} \tag{M.40a}$$

$$\sigma^2_{\mathrm{prior}} \equiv \mu^2_{\mathrm{prior}} \left[e^{s^2_{\mathrm{prior}}} - 1\right]\,. \tag{M.40b}$$

This analysis tells us that the distribution $P(f_{\mathrm{lin}}|w_i, x_i, \lambda_{\mathrm{tar},i})$ is Gaussian in $e^{\lambda_{\mathrm{tar},i}}$. To determine the learning rules, all we have to do is insert Eq. (M.37) into Eq. (M.31), perform an integral, take the log, compute the first two derivatives, and evaluate them at $m_i$ (see Eq. M.29). These steps, which are performed in Supplementary Math Note, Sec. S3.1, are not completely straightforward, as various approximations must be made. However, from a conceptual point of view the approximations do not add much. Thus, here we simply give the results.

### Linear feedback, $f = f_{\mathrm{lin}}$

The Bayesian update rules are

$$\Delta m_i = \left(\frac{s^2_i \mu_i}{\sigma^2_{\delta 0}}\right) x_i f_{\mathrm{lin}} - \frac{1}{\tau}\left(m_i - m_{\mathrm{prior}}\right) \tag{M.41a}$$

$$\Delta s^2_i = -\left(\frac{s^4_i \mu^2_i}{\sigma^2_{\delta 0}}\right) x^2_i - \frac{2}{\tau}\left(s^2_i - s^2_{\mathrm{prior}}\right)\,. \tag{M.41b}$$

For classical learning, we use the delta rule (main text, Eq. 3),

$$\Delta w_i = \eta \, x_i f_{\text{lin}} \,. \tag{M.42}$$

Note that we are not including weight drift in the classical learning rate (both here and below), as weight drift was derived using Bayesian analysis, and has no classical counterpart.

**Cerebellar feedback,** $f = f_{\mathbf{cb}} = \Theta(f_{\mathbf{lin}} - \theta)$

The Bayesian update rules are

$$\Delta m_i = \left(\frac{s_i^2 \mu_i}{\sigma_{\delta 0}^2}\right) x_i \sigma_{\delta 0} (2 f_{\text{cb}} - 1) \, \frac{\mathcal{N}(\theta_{\text{cb}})}{\Phi\big(\theta_{\text{cb}}\big)} - \frac{1}{\tau}(m_i - m_{\text{prior}}) \tag{M.43a}$$

$$\Delta s_i^2 = -\left(\frac{s_i^4 \mu_i^2}{\sigma_{\delta 0}^2}\right) x_i^2 \, \frac{\mathcal{N}(\theta_{\text{cb}})}{\Phi\big(\theta_{\text{cb}}\big)} \left[\theta_{\text{cb}} + \frac{\mathcal{N}\big(\theta_{\text{cb}}\big)}{\Phi\big(\theta_{\text{cb}}\big)}\right] - \frac{2}{\tau}\left(s_i^2 - s_{\text{prior}}^2\right) \tag{M.43b}$$

where $\Phi$ and (in a slight abuse of notation) $\mathcal{N}$ are the cumulative normal and normal functions, respectively,

$$\Phi(z) \equiv \int_{-\infty}^{z} du \, \frac{e^{-u^2/2}}{(2\pi)^{1/2}} \tag{M.44a}$$

$$\mathcal{N}(z) \equiv \frac{e^{-z^2/2}}{(2\pi)^{1/2}} \,, \tag{M.44b}$$

and $\theta_{\text{cb}}$ is given in terms of the threshold, $\theta$, as

$$\theta_{\text{cb}} \equiv (1 - 2 f_{\text{cb}}) \, \frac{\theta}{\sigma_{\delta 0}} \,. \tag{M.45}$$

For classical learning, we absorb most of the prefactor in the above mean update into a fixed learning rate,

$$\Delta w_i = \eta (2 f_{\text{cb}} - 1) x_i \, \frac{\mathcal{N}(\theta_{\text{cb}})}{\Phi\big(\theta_{\text{cb}}\big)} \,. \tag{M.46}$$

**Reinforcement learning,** $f = f_{\mathbf{rl}} = -|f_{\mathbf{lin}}|$

The Bayesian update rules are

$$\Delta m_i = \left(\frac{s_i^2 \mu_i}{\sigma_\delta^2}\right) \left(\frac{f_{\text{rl}}^2}{\sigma_\delta^2} - 1\right) x_i^2 (\mu_i - w_i) - \frac{1}{\tau}(m_i - m_{\text{prior}}) \tag{M.47a}$$

$$\Delta s_i^2 = -\left(\frac{s_i^4 \mu_i^2}{\sigma_\delta^2}\right) \left(1 - \frac{f_{\text{rl}}^2}{\sigma_\delta^2}\right) x_i^2 - \frac{2}{\tau}\left(s_i^2 - s_{\text{prior}}^2\right) . \tag{M.47b}$$

This learning rule appears non-local, as it depends on $\sigma_\delta^2$, which in turn depends on all the synapses (Eq. M.36). However, we make it local by changing the feedback signal to $(1 - f_{\rm rl}^2/\sigma_\delta^2)/\sigma_\delta^2$. For classical learning, we again absorb most of the prefactor into the learning rate,

$$\Delta w_i = \eta x_i \big( f_{\rm rl} \tanh\big( (\mu_i - w_i) x_i f_{\rm rl}/\sigma_\delta^2 \big) - (\mu_i - w_i) x_i \big) \,. \tag{M.48}$$

Note that tanh appears in the classical, but not Bayesian, learning rules. That is because for the Bayesian learning rules we made the approximation $\tanh\big((\mu_i - w_i)x_i f_{\rm rl}/\sigma_\delta^2\big) \approx (\mu_i - w_i)x_i f_{\rm rl}/\sigma_\delta^2$. This approximation, however, made the classical learning rule unstable.

### M2.3 Recurrent neural network learning rules

So far we have focused on single neurons. Here we generalize to the more realistic scenario in which the output weights of a recurrent network are trained to produce a time-dependent target function. We will assume that the network, which contains $N$ neurons, evolves according to

$$\tau_m \frac{dv_i}{dt} = -v_i + \sum_{j=1}^{N} J_{ij} x_j + A_i V(t) + I_i(t) \tag{M.49a}$$

$$x_j = \tanh(v_j) \tag{M.49b}$$

$$V(t) = \sum_{j=1}^{N} w_j x_j \,. \tag{M.49c}$$

We interpret $v_i$ as the membrane potential and $x_j$ as the firing rate relative to baseline. The recurrent weights, $J_{ij}$, and feedback weights, $A_i$, are fixed. Parameters of the network, and details of the simulations, are given in Supplementary Math Note, Sec. S5.

The goal of the network is to minimize the distance between $V(t)$ and some target function, denoted $V_{\rm tar}(t)$; that is, to minimize the error, $\delta(t)$, defined, as in Eq. (2), to be

$$\delta(t) \equiv V_{\rm tar}(t) - V(t) \,. \tag{M.50}$$

As with single neurons, we take a Bayesian approach. There are, however, two important differences. The first is that we do not know the target weights (we do not specify them; instead they must be learned). We assume, though, that target weights exist, which means we can write

$$\delta(t) = \sum_j \big( w_{{\rm tar},j}(t) - w_j(t) \big) x_j(t) \,. \tag{M.51}$$

The second difference is that the feedback signal, $\delta(t)$, is a continuous function of time. Consequently, information at times $t$ and $t + dt$ is largely redundant. To deal with this redundancy we make several approximations. First, rather than updating the weights continuously, we update them at times separated by $\Delta t$. (In a slight abuse of notation, here $\Delta t$ does not have the same numerical value as in the single neuron update rules.) Bayes' theorem, Eq. (M.20), then becomes

$$P(w_{\text{tar},i}|\mathcal{D}_i) \propto P(d_i|w_{\text{tar},i}, \mathcal{D}_i(t-\Delta t))\, P(w_{\text{tar},i}|\mathcal{D}_i(t-\Delta t)) \tag{M.52}$$

where, as in the single neuron case, the data for synapse $i$ is the presynaptic input, $x_i$, the actual weight, $w_i$, and the error signal, $\delta$. To derive this expression we made two simplifications: we did not add noise to the error signal, so the synapses see $\delta$ rather than $f_{\text{lin}}$, and we did not enforce Dale's law, so the weights can change sign. Because of the latter simplification, we let the weights, rather than the log weights, have a Gaussian distribution; that is why Eq. (M.52) is written in terms of $w_{\text{tar},i}$ rather than $\lambda_{\text{tar},i}$.

In one respect the analysis is simpler than it was for single neurons: because the target weights do not evolve over time (see comments at the end of Sec. M1.1), we can avoid the integral in Eq. (M.21). However, in another respect it is more complicated: as just discussed, the likelihood (the first term on the right hand side of Eq. M.52) depends on past data. An exact treatment in this regime is beyond the scope of this work. What we do instead is choose the time step, $\Delta t$, so it is much larger than the correlation time of $\delta(t)$. This allows us to drop the dependence on $\mathcal{D}_i(t-\Delta t)$ in the likelihood, giving us

$$P(d_i|w_{\text{tar},i}, \mathcal{D}_i(t-\Delta t)) \approx P(d_i|w_{\text{tar},i}) \propto P(\delta|x_i, w_i, w_{\text{tar},i}) \tag{M.53}$$

where, as in Eq. (M.30), we used the fact that without an error signal, $x_i$ and $w_i$ do not provide any information about $w_{\text{tar},i}$.

While this gives us a very good approximation to the likelihood if $\Delta t$ is large, large $\Delta t$ means that updates would be made very rarely, and so learning would be slow. We thus make a second approximation, which is to optimize our learning rule (via numerical simulation, as discussed below) with respect to $\Delta t$. This gives us approximate Bayesian update rules, which presumably could be improved upon. However, as we will see, the approximate update rules already outperform the classical ones by an order of magnitude. Thus, any improvement would only make the case for Bayesian plasticity stronger.

To find an expression for $P(\delta|x_i, w_i, w_{\text{tar},i})$, we again write $\delta$ as in Eq. (M.32) (but without noise, so $\xi_\delta = 0$, which reduces $\sigma_0^2$; see Eq. M.33). Now, however, we are interested in the log likelihood with respect to the target weights, $w_{\text{tar},i}$, rather than the log of the target weights, $\lambda_{\text{tar},i}$ (as mentioned above). Thus, the distribution over $\delta$ simplifies relative to Eq. (M.37),

$$\delta|w_i, x_i, w_{\text{tar},i} \sim \mathcal{N}\left((w_{\text{tar},i} - w_i)x_i, \sigma_\delta^2\right) . \tag{M.54}$$

As above, we made the approximation $\sigma_{\delta,i}^2 \approx \sigma_\delta^2$; see Eq. (M.36). It is now straightforward to write down the log likelihood,

$$L(w_{\text{tar},i}) = -\frac{(\delta - (w_{\text{tar},i} - w_i)\, x_i)^2}{2\sigma_\delta^2} + \text{const.} \tag{M.55}$$

The first and second derivatives evaluated at $w_{\text{tar},i} = \mu_i$ are

$$L'(\mu_i) = \frac{(\delta - (\mu_i - w_i)\, x_i) x_i}{\sigma_\delta^2} \approx \frac{\delta x_i}{\sigma_\delta^2} \tag{M.56a}$$

$$L''(\mu_i) = -\frac{x_i^2}{\sigma_\delta^2}\,. \tag{M.56b}$$

(We are justified in dropping the term $(\mu_i - w_i) x_i$ because it is a factor of $\sqrt{n}$ smaller than $\delta$. That follows because $\sigma_\delta^2$, which is the variance of $\delta$, is $\mathcal{O}(n)$; see Eq. M.36.) Inserting these expressions into Eq. (M.29) (with $\tau$ taken to $\infty$ because, as discussed in Sec. M1.1, we are assuming the target weights do not drift over time), we have

$$\Delta\mu_i = \frac{\sigma_i^2}{\sigma_\delta^2}\,\delta x_i \tag{M.57a}$$

$$\Delta\sigma_i^2 = -\frac{\sigma_i^4}{\sigma_\delta^2}\,x_i^2\,, \tag{M.57b}$$

where $\Delta\mu_i = \mu_i(t + \Delta t) - \mu_i(t)$ and similarly for $\Delta\sigma_i^2$.

Primarily for convenience, we make a third approximation, which is to update the weights continuously rather than at discrete points separated by $\Delta t$. To do that we simply make the approximation $\Delta\mu_i \approx \Delta t\, d\mu_i/dt$, and similarly for $\Delta\sigma_i^2$. This allows us to turn the update rules into ordinary differential equations,

$$\frac{d\mu_i}{dt} = \frac{1}{\Delta t}\frac{\sigma_i^2}{\sigma_\delta^2}\,\delta x_i \tag{M.58a}$$

$$\frac{d\sigma_i^2}{dt} = -\frac{1}{\Delta t}\frac{\sigma_i^4}{\sigma_\delta^2}\,x_i^2\,. \tag{M.58b}$$

Then, defining

$$\eta_i \equiv \frac{\sigma_i^2}{\sigma_\delta^2 \Delta t}\,, \tag{M.59}$$

inserting this into into Eq. (M.58) and, in our fourth approximation, ignoring the time dependence in $\sigma_\delta^2$, those equations simplify to

$$\frac{d\mu_i}{dt} = \eta_i\,\delta x_i \tag{M.60a}$$

$$\frac{d\eta_i}{dt} = -\eta_i^2\,x_i^2\,. \tag{M.60b}$$

Optimizing over $\Delta t$ corresponds to optimizing over the initial value of $\eta_i$, which we assumed is the same for all $i$. This optimization is done via numerical simulations.

For the classical learning rules, we drop Eq. (M.60b) and fix $\eta_i$ to the same value for all synapses.

Further methods details can be found in the attached Life Sciences Reporting Summary.

# Code Availability

Code and data are available for download at:
https://github.com/Jegmi/the-bayesian-synapse/releases/tag/v2

# Data Availability

# References

[1] Y. Loewenstein, A. Kuras, and S. Rumpel. "Multiplicative Dynamics Underlie the Emergence of the Log-Normal Distribution of Spine Sizes in the Neocortex In Vivo". In: *Journal of Neuroscience* 31.26 (2011), pp. 9481–9488.

[2] M. Matsuzaki et al. "Structural basis of long-term potentiation in single dendritic spines". In: *Nature* 429.6993 (2004), pp. 761–766.

[3] S. Song et al. "Highly nonrandom features of synaptic connectivity in local cortical circuits". In: *PLoS Biology* 3.3 (2005), e68.

[4] D. H. O'Connor et al. "Neural activity in barrel cortex underlying vibrissa-based object localization in mice". In: *Neuron* 67.6 (2010), pp. 1048–1061.

[5] K. Mizuseki and G. Buzsáki. "Preconfigured, Skewed Distribution of Firing Rates in the Hippocampus and Entorhinal Cortex". In: *Cell Reports* 4.5 (2013), pp. 1010–1021.

[6] T. P. Minka. "A family of algorithms for approximate Bayesian inference". PhD thesis. Massachusetts Institute of Technology, 2001.

# Supplementary Math Note

Here we provide a complete derivation of the Bayesian and classical learning rules for linear, cerebellar and reinforcement learning, and derive the simplified learning rules used in Eq. (4) of the main text (Sec. S3), derive the relationship between variability and firing rate (Sec. S4), provide details of the simulations (Sec. S5, which includes tables containing the parameters used in the simulations), discuss how we chose the parameters of our model (Sec. S6), provide a normative explanation for the Synaptic Sampling hypothesis (Sec. S7), and explore the robustness of our model (Sec. S8).

## S3 Derivation of the learning rules

The learning rules derived in Methods, Eq. (M.29), all depend on the likelihood of the data given the target weights. Here we compute the likelihood for our three single neuron feedback signals, and use those to write down the learning rules. The learning rules for the recurrent neural network were derived in Methods, Sec. M2.3.

### S3.1 Single neuron learning rules in terms of the log likelihood

Our starting point is the likelihood given in Methods, Eq. (M.31); the corresponding log likelihood is

$$L(\lambda_{\text{tar},i}) = \log \int df_{\text{lin}}\, P(f|f_{\text{lin}})\, P(f_{\text{lin}}|x_i, w_i, \lambda_{\text{tar},i}) \tag{S.61}$$

where $f$ is either $f_{\text{lin}}$, $f_{\text{cb}}$ or $f_{\text{rl}}$, all of which are deterministic functions of $f_{\text{lin}}$ (see main text).

The second term inside the integral, $P(f_{\text{lin}}|x_i, w_i, \lambda_{\text{tar},i})$, is given in Methods, Eq. (M.37). Although that distribution is Gaussian in $f_{\text{lin}}$, its mean depends on $e^{\lambda_{\text{tar},i}}$, so as a function of $\lambda_{\text{tar},i}$ it is somewhat unwieldy. We thus use statistical linearisation [1] to simplify it: we approximate $\pm e^{\lambda_{\text{tar},i}}$ with a straight line, $a(\lambda_{\text{tar},i} - m_i) + b$, with $a$ and $b$ chosen to minimize the expected mean squared error between $\pm e^{\lambda_{\text{tar},i}}$ and $a(\lambda_{\text{tar},i} - m_i) + b$, where the expectation is with respect to the posterior on the previous time step, $P(\lambda_{\text{tar},i}|\mathcal{D}(t-1))$. The solution is especially simple, $a = b = \mu_i$, which gives

$$\pm e^{\lambda_{\text{tar},i}} \approx \mu_i \left(1 + \lambda_{\text{tar},i} - m_i\right) \tag{S.62}$$

(recall that $\mu_i$ is the expected value of $w_{\text{tar},i}$; see Methods, Eq. (M.13a)). Inserting this into Methods, Eq. (M.37), we have

$$\begin{aligned} L_{\text{lin}}(\lambda_{\text{tar},i}) &\equiv \log P(f_{\text{lin}}|x_i, w_i, \lambda_{\text{tar},i}) \\ &= -\frac{\left(f_{\text{lin}} - (\mu_i - w_i)\, x_i - \mu_i x_i \left(\lambda_{\text{tar},i} - m_i\right)\right)^2}{2\sigma_\delta^2} + \text{const} \end{aligned} \tag{S.63}$$

where $\sigma_\delta^2$ is given in Methods, Eq. (M.36). Note that equality in this expression (and many that follow) is shorthand for equality under the assumption that $P(f_{\text{lin}}|x_i, w_i, \lambda_{\text{tar},i})$ is given by Eq. (S.63). Inserting Eq. (S.63) into the expression for the log likelihood, Eq. (S.61), we arrive at

$$L(\lambda_{\text{tar},i}) = \log \int df_{\text{lin}}\, P(f|f_{\text{lin}})\, e^{L_{\text{lin}}(\lambda_{\text{tar},i})} + \text{const}. \tag{S.64}$$

To write down an explicit expression for $\sigma_\delta^2$ – which we will need below – we use the fact that $\sigma_i^2$ is the variance of $e^{\lambda_{\text{tar},i}}$. Then, because the synapses are independent, $\sigma_\delta^2$ can be written as combination of the uncertainty about the target weight, $w_{\text{tar},i}$, and the variability in the actual weight due to noise,

$$\sigma_\delta^2 = \sum_j \left(\sigma_j^2 + \text{Var}\left[w_j\right]\right) x_j^2 + \sigma_0^2 \tag{S.65}$$

where, recall, $\sigma_0^2$ is the combined variance of the noise in the membrane potential and the feedback signal (Methods, Eq. (M.33)). What we use for the variance of $w_j$ depends on whether we are sampling (for which $\text{Var}\left[w_i\right] = \sigma_i^2$), or using a variance proportional to the mean, (for which $\text{Var}\left[w_i\right] = k\mu_i$); see Methods, Eq. (M.39).

Derivatives of the log likelihood, $L(\lambda_{\text{tar},i})$ – the main ingredients in the learning rule, Methods, Eq. (M.29) – can be expressed as derivatives of $L_{\text{lin}}$; as is straightforward to show,

$$L'(\lambda_{\text{tar},i}) = \mathbb{E}_{\text{lin}}\left[L'_{\text{lin}}(\lambda_{\text{tar},i})\right] \tag{S.66a}$$

$$L''(\lambda_{\text{tar},i}) = \mathbb{E}_{\text{lin}}\left[L''_{\text{lin}}(\lambda_{\text{tar},i})\right] + \text{Var}_{\text{lin}}\left[L'_{\text{lin}}(\lambda_{\text{tar},i})\right] \tag{S.66b}$$

where the expectation and variance are with respect to the distribution

$$P(f_{\text{lin}}|f, x_i, w_i, \lambda_{\text{tar},i}) = \frac{P(f|f_{\text{lin}})\, e^{L_{\text{lin}}(\lambda_{\text{tar},i})}}{\int df'_{\text{lin}} P(f|f'_{\text{lin}})\, e^{L_{\text{lin}}(\lambda_{\text{tar},i})}}\,. \tag{S.67}$$

(Recall that $L_{\text{lin}}(\lambda_{\text{tar},i})$ depends on $f_{\text{lin}}$; see Eq. (S.63).) Using Eq. (S.63) for $L_{\text{lin}}(\lambda_{\text{tar},i})$ the derivatives are straightforward; combining those with Methods, Eq. (M.29) (for which we need to evaluate the derivatives at $\lambda_{\text{tar},i} = m_i$), we have, after a small amount of algebra,

$$\Delta m_i = \left(\frac{s_i^2 \mu_i}{\sigma_\delta^2}\right) x_i \big(\mathbb{E}_{\text{lin}}\left[f_{\text{lin}}\right] + x_i\left(w_i - \mu_i\right)\big) - \frac{m_i - m_{\text{prior}}}{\tau} \tag{S.68a}$$

$$\Delta s_i^2 = -\left(\frac{s_i^4 \mu_i^2}{\sigma_\delta^2}\right) x_i^2 \left(\frac{\sigma_\delta^2 - \text{Var}_{\text{lin}}\left[f_{\text{lin}}\right]}{\sigma_\delta^2}\right) - \frac{2\left(s_i^2 - s_{\text{prior}}^2\right)}{\tau}\,. \tag{S.68b}$$

In the next section, we use these expressions to derive the explicit learning rules for our three single neuron feedback signals, $f_{\text{lin}}$, $f_{\text{cb}}$ and $f_{\text{rl}}$. Note that because $\lambda_{\text{tar},i}$ is evaluated at $\lambda_{\text{tar},i} = m_i$, the expression for $L_{\text{lin}}(\lambda_{\text{tar},i})$, Eq. (S.63), simplifies considerably.

In Fig. S.6 we show a dependency graph which describes how each variable is generated. This is a graphical model — a compact method for describing dependencies among random variables. This graphical model has the unusual feature that the results of inference at one time step influence the data at subsequent time steps.

## S3.2 Single neuron learning rules for our three feedback signals

To derive explicit learning rules for our three feedback signals, we just need to compute the mean and variance of $f_{\text{lin}}$, and then insert those into Eq. (S.68). Those calculations, which are mainly straightforward, follow.

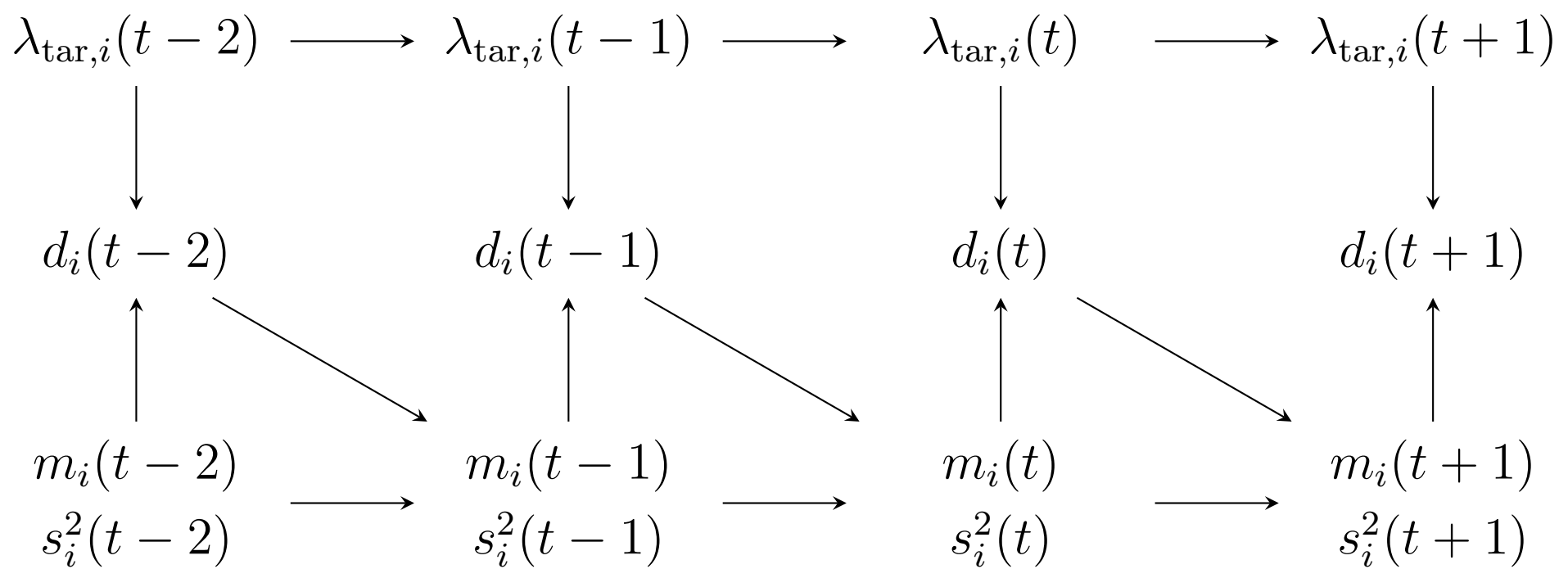

Figure S.6: The graphical model describing the dependencies in our simulations. The log of the target weight, $\lambda_{\text{tar},i}(t)$, evolves independently of all other variables, under the Ornstein-Uhlenbeck (OU) process described in Methods, Eq. (M.11). The data, $d_i(t)$, which consists of the presynaptic input, $x_i(t)$, the actual weight, $w_i(t)$, and a signal that provides information about the target weights (Methods, Eq. (M.18)), depends on both the target weight, $\lambda_{\text{tar},i}(t)$, and on past inference, $m_i(t)$ and $s_i^2(t)$. The mean and uncertainty at time $t$, $m_i(t)$ and $s_i^2(t)$, depend on the mean and uncertainty at the previous time step, $m_i(t-1)$ and $s_i^2(t-1)$, and also on past data, $d_i(t-1)$, through the learning rules (Methods, Eq. (M.29)).

### Linear feedback, $f = f_{\mathbf{lin}}$

For linear feedback, $\mathbb{E}_{\text{lin}}\left[f_{\text{lin}}\right] = f$ and $\text{Var}_{\text{lin}}\left[f_{\text{lin}}\right] = 0$. Thus, the update rules for the mean and variance of the log weights, Eq. (S.68), become

$$\Delta m_i = \left(\frac{s_i^2 \mu_i}{\sigma_\delta^2}\right) x_i \left(f_{\text{lin}} + x_i(w_i - \mu_i)\right) - \frac{1}{\tau}\left(m_i - m_{\text{prior}}\right), \tag{S.69a}$$

$$\Delta s_i^2 = -\left(\frac{s_i^4 \mu_i^2}{\sigma_\delta^2}\right) x_i^2 - \frac{2}{\tau}\left(s_i^2 - s_{\text{prior}}^2\right). \tag{S.69b}$$

While these rules are easily updated on a digital computer, they are somewhat problematic for real synapses, since they are non-local: $\sigma_\delta^2$ depends on all the synapses (see Eq. (S.65)). To remedy this, we replace $\sigma_\delta^2$ with a constant, $\sigma_{\delta 0}^2$, the the average value of $\sigma_\delta^2$ under the prior (see Methods, Eq. (M.38)). In addition, as in Methods (see discussion following Eq. (M.56)), we drop the term $x_i(w_i - \mu_i)$ in Eq. (S.69a), as it is small compared to $f_{\text{lin}}$. The resulting learning rules, along with their classical counterparts, are given in Methods, Eqs. (M.41) and (M.42).

### Cerebellar feedback, $f = f_{\mathbf{cb}} = \mathbf{sign}(f_{\mathbf{lin}} - \theta)$

For cerebellar feedback, $P(f_{\text{cb}}|f_{\text{lin}})$ may be written

$$P(f_{\text{cb}}|f_{\text{lin}}) = f_{\text{cb}}\Theta(f_{\text{lin}} - \theta) + (1 - f_{\text{cb}})\Theta(\theta - f_{\text{lin}}) \tag{S.70}$$

where $f_{\rm cb}$ can take on only the values 0 and 1 and, as usual, $\Theta$ is the Heaviside step function. This expression tells us that if $f_{\rm cb} = 1$ then $f_{\rm lin} \geq \theta$, and if $f_{\rm cb} = 0$ then $f_{\rm lin} < \theta$. Consequently, $P(f_{\rm lin}|f_{\rm cb}, x_i, w_i, m_i)$, Eq. (S.67), is a truncated Gaussian whose mean and variance can be computed in terms of the cumulative normal function.

To derive the learning rules, it is easiest to compute $L(\lambda_{{\rm tar},i})$ directly from Eq. (S.64) and then take derivatives with respect to $\lambda_{{\rm tar},i}$, rather than using Eq. (S.68). To this end, we insert Eq. (S.70) into Eq. (S.64); then, using Eq. (S.63) for $L_{\rm lin}(\lambda_{{\rm tar},i})$, we have

$$L(\lambda_{{\rm tar},i}) = \log \Bigg[ \int df_{\rm lin} \left( f_{\rm cb}\Theta(f_{\rm lin} - \theta) + (1 - f_{\rm cb})\Theta(\theta - f_{\rm lin}) \right) \times \exp\left( -\left(f_{\rm lin} - (\mu_i - w_i)x_i - \mu_i x_i(\lambda_{{\rm tar},i} - m_i)\right)^2 / 2\sigma_\delta^2 \right) \Bigg] + \text{const.} \tag{S.71}$$

The integral is straightforward, and we arrive at

$$L(\lambda_{{\rm tar},i}) = \log \left[ \Phi\left( (2f_{\rm cb} - 1) \left[ \frac{(\mu_i - w_i)x_i + \mu_i x_i(\lambda_{{\rm tar},i} - m_i)}{\sigma_\delta^2} - \theta \right] \right) \right] + \text{const} \tag{S.72}$$

where $\Phi$ is the cumulative normal function, defined in Methods, Eq. (M.44a). Taking the first and second derivatives with respect to $\lambda_{{\rm tar},i}$ and evaluating them at $\lambda_{{\rm tar},i} = m_i$ gives us

$$L'(m_i) = \frac{\mu_i x_i}{\sigma_\delta} (2f_{\rm cb} - 1) \frac{\mathcal{N}(\tilde\theta_{\rm cb})}{\Phi(\tilde\theta_{\rm cb})} \tag{S.73a}$$

$$L''(m_i) = -\frac{\mu_i^2 x_i^2}{\sigma_\delta^2} \frac{\mathcal{N}(\tilde\theta_{\rm cb})}{\Phi(\tilde\theta_{\rm cb})} \left[ \tilde\theta_{\rm cb} + \frac{\mathcal{N}(\tilde\theta_{\rm cb})}{\Phi(\tilde\theta_{\rm cb})} \right] \tag{S.73b}$$

where $\tilde\theta_{\rm cb}$ is given by

$$\tilde\theta_{\rm cb} \equiv (2f_{\rm cb} - 1) \frac{(\mu_i - w_i)x_i - \theta}{\sigma_\delta} \tag{S.74}$$

and $\mathcal{N}$ is the standard normal function (Methods, Eq. (M.44b)). Inserting these expressions into Methods, Eq (M.29), leads to

$$\Delta m_i = \left( \frac{s_i^2 \mu_i}{\sigma_\delta^2} \right) x_i \sigma_\delta (2f_{\rm cb} - 1) \frac{\mathcal{N}(\tilde\theta_{\rm cb})}{\Phi\big(\tilde\theta_{\rm cb}\big)} - \frac{1}{\tau}(m_i - m_{\rm prior}) \tag{S.75a}$$

$$\Delta s_i^2 = -\left( \frac{s_i^4 \mu_i^2}{\sigma_\delta^2} \right) x_i^2 \frac{\mathcal{N}(\tilde\theta_{\rm cb})}{\Phi\big(\tilde\theta_{\rm cb}\big)} \left[ \tilde\theta_{\rm cb} + \frac{\mathcal{N}\big(\tilde\theta_{\rm cb}\big)}{\Phi\big(\tilde\theta_{\rm cb}\big)} \right] - \frac{2}{\tau} \left( s_i^2 - s_{\rm prior}^2 \right). \tag{S.75b}$$

As with linear feedback, to ensure that the learning rule is local, we replace $\sigma_\delta^2$ with $\sigma_{\delta 0}^2$. In addition, we drop the contribution of $(\mu_i - w_i)x_i$ to $\tilde\theta_{\rm cb}$ (Eq. (S.74)). That follows from the same reasoning we used for the linear feedback: $\theta$ is on the order of $\sigma_\delta$, which is much larger than $(\mu_i - w_i)x_i$. We thus define $\theta_{\rm cb}$ to be $\tilde\theta_{\rm cb}$ but without the term $(\mu_i - w_i)x_i$ and with $\sigma_\delta$ replaced with $\sigma_{\delta 0}$, The resulting learning rules, along with their classical counterparts, are given in Methods, Eqs. (M.43) and (M.46).

**Reinforcement learning, $f = f_{\mathrm{rl}} = -|f_{\mathrm{lin}}|$**

For reinforcement learning,

$$P\left(f_{\mathrm{rl}}|f_{\mathrm{lin}}\right) = \delta\left(f_{\mathrm{rl}} + |f_{\mathrm{lin}}|\right). \tag{S.76}$$

Combining this with Eq. (S.63) for $L(\lambda_{\mathrm{tar},i})$ and using Eq. (S.67), we have

$$P(f_{\mathrm{lin}}|f_{\mathrm{rl}}, x_i, w_i, m_i) \propto \delta(f_{\mathrm{rl}} + |f_{\mathrm{lin}}|) \exp\left(f_{\mathrm{lin}}(\mu_i - w_i)x_i/\sigma_\delta^2\right) \tag{S.77}$$

where we evaluated $\lambda_{\mathrm{tar},i}$ at $\lambda_{\mathrm{tar},i} = m_i$ and used the fact that, because of the delta-function, $f_{\mathrm{lin}}^2 = f_{\mathrm{rl}}^2$. Consequently, the expectations needed for the update rule, Eq. (S.68), are

$$\mathbb{E}\left[f_{\mathrm{lin}}|f_{\mathrm{rl}}, x_i, w_i, m_i\right] = f_{\mathrm{rl}} \tanh\left((\mu_i - w_i)x_i f_{\mathrm{rl}}/\sigma_\delta^2\right) \tag{S.78a}$$

$$\mathrm{Var}\left[f_{\mathrm{lin}}|f_{\mathrm{rl}}, x_i, w_i, m_i\right] = \frac{f_{\mathrm{rl}}^2}{\cosh^2\left((\mu_i - w_i)x_i f_{\mathrm{rl}}/\sigma_\delta^2\right)}. \tag{S.78b}$$

Inserting these into Eq. (S.68) yields

$$\Delta m_i = \left(\frac{s_i^2\mu_i}{\sigma_\delta^2}\right) x_i\left(f_{\mathrm{rl}} \tanh\left((\mu_i - w_i)x_i f_{\mathrm{rl}}/\sigma_\delta^2\right) - (\mu_i - w_i)x_i\right) - \frac{1}{\tau}(m_i - m_{\mathrm{prior}}) \tag{S.79a}$$

$$\Delta s_i^2 = -\left(\frac{s_i^4\mu_i^2}{\sigma_\delta^2}\right) x_i^2 \left(1 - \frac{f_{\mathrm{rl}}^2/\sigma_\delta^2}{\cosh^2\left((\mu_i - w_i)x_i f_{\mathrm{rl}}/\sigma_\delta^2\right)}\right) - \frac{2}{\tau}\left(s_i^2 - s_{\mathrm{prior}}^2\right). \tag{S.79b}$$

Not surprisingly, for reinforcement learning the synapse must explore: if it sets the weight, $w_i$, to $\mu_i$ (the mean value of $w_{\mathrm{tar},i}$), $m_i$ will relax to the prior. Sampling in this case is critical.

Because $f_{\mathrm{rl}}$ and $\sigma_\delta$ are both large, we can make the approximations $\tanh z \approx z$ and $\cosh^2 z \approx 1$, yielding simplified learning rules. Those learning rules, along with their classical counterparts (for which do not make the approximation $\tanh z \approx z$, as it leads to an instability), are given in Methods, Eqs. (M.47) and (M.48).

### S3.3 Simplifying the single neuron learning rules

In our simulations we use the update rules derived above (and given explicitly in Methods, Sec. M2.2). However, for illustrative purposes, in the main text, Eq. (4), we gave simplified learning rules relevant to linear feedback, $f = f_{\mathrm{lin}} = \delta + \xi_\delta$. Here we derive those simplified rules. The derivation involves rather severe approximations; we make them only to illustrate the essence of the learning rules in the simplest possible setting.

For this analysis we consider the small noise regime, $s_i^2 \ll 1$. In that regime, Methods, Eq. (M.13b) becomes

$$\sigma_i^2 \approx \mu_i^2 s_i^2. \tag{S.80}$$

Combining this with Methods, Eq. (M.13a), we have

$$\Delta\mu_i = \mu_i \left(\Delta m_i + \tfrac{1}{2}\Delta s_i^2\right) \tag{S.81a}$$

$$\Delta\sigma_i^2 = 2\mu_i^2 s_i^2 \left(\Delta m_i + \tfrac{1}{2}\Delta s_i^2\right) + \mu_i^2 \Delta s_i^2 \,. \tag{S.81b}$$

Because $\Delta s_i^2$ is a factor of $s_i^2$ smaller than $\Delta m_i$ (see Eq. (S.69)), our small noise approximation lets us drop the term proportional to $\Delta s_i^2$ in the first expression and the term proportional to $s_i^2$ in the second expression. Consequently, the update rules for $\mu_i$ and $\sigma_i^2$ become

$$\Delta\mu_i \approx \mu_i \Delta m_i \tag{S.82a}$$

$$\Delta\sigma_i^2 \approx \mu_i^2 \Delta s_i^2 \,. \tag{S.82b}$$

Inserting these into Eq. (S.69) and using the approximation given in Eq. (S.80), we arrive at

$$\Delta\mu_i \approx \frac{\sigma_i^2}{\sigma_\delta^2} x_i f_{\text{lin}} - \frac{\mu_i}{\tau}\left(m_i - m_{\text{prior}}\right) \tag{S.83a}$$

$$\Delta\sigma_i^2 \approx -\frac{\sigma_i^4}{\sigma_\delta^2} x_i^2 - \frac{2\mu_i^2}{\tau}\left(s_i^2 - s_{\text{prior}}^2\right) \,. \tag{S.83b}$$

As in Methods, we neglected the term $x_i(w_i - \mu_i)$ because it is small compared to $f_{\text{lin}}$ (see comments following Eq. (M.56)).

To show that the contribution from the prior is approximately the form given in the main text, Eq. (4a), we use Methods, Eq. (M.13a) to write

$$\mu_i - \mu_{\text{prior}} = \mu_i \left(1 - e^{-\left(m_i - m_{\text{prior}}\right) - \left(s_i^2 - s_{\text{prior}}^2\right)/2}\right) . \tag{S.84}$$

Taylor expanding the exponent and neglecting both $s_i^2$ and $s_{\text{prior}}^2$, we arrive at

$$\mu_i - \mu_{\text{prior}} \approx \mu_i \left(m_i - m_{\text{prior}}\right) \,. \tag{S.85}$$

Similarly, we use our standard approximation for the variance,

$$\sigma_i^2 - \sigma_{\text{prior}}^2 \approx s_i^2 \mu_i^2 - s_{\text{prior}}^2 \mu_{\text{prior}}^2 \approx \mu_i^2 (s_i^2 - s_{\text{prior}}^2) \tag{S.86}$$

where the last expression follows, approximately, because $\mathbb{E}\left[\mu_i\right] = \mu_{\text{prior}}$.

The learning rules given in Eq. (S.83) imply that the change in the mean weight, $\Delta\mu_i$, is proportional to the variance, $\sigma_i^2$. Thus, the relative change in the mean weight, $\Delta\mu_i/\mu_i$, is proportional to the variance divided by the mean, which is the normalized uncertainty. Under sampling, the latter should be equal to the observed normalized uncertainty. This is a very general feature: under the approximations given in Eq. (S.82), for all of our learning rules (Eqs. (M.41), (M.43) and (M.47)), $\Delta\mu_i/\mu_i \propto \sigma_i^2/\mu_i$. We see this in our simulations; see Fig. S.7.

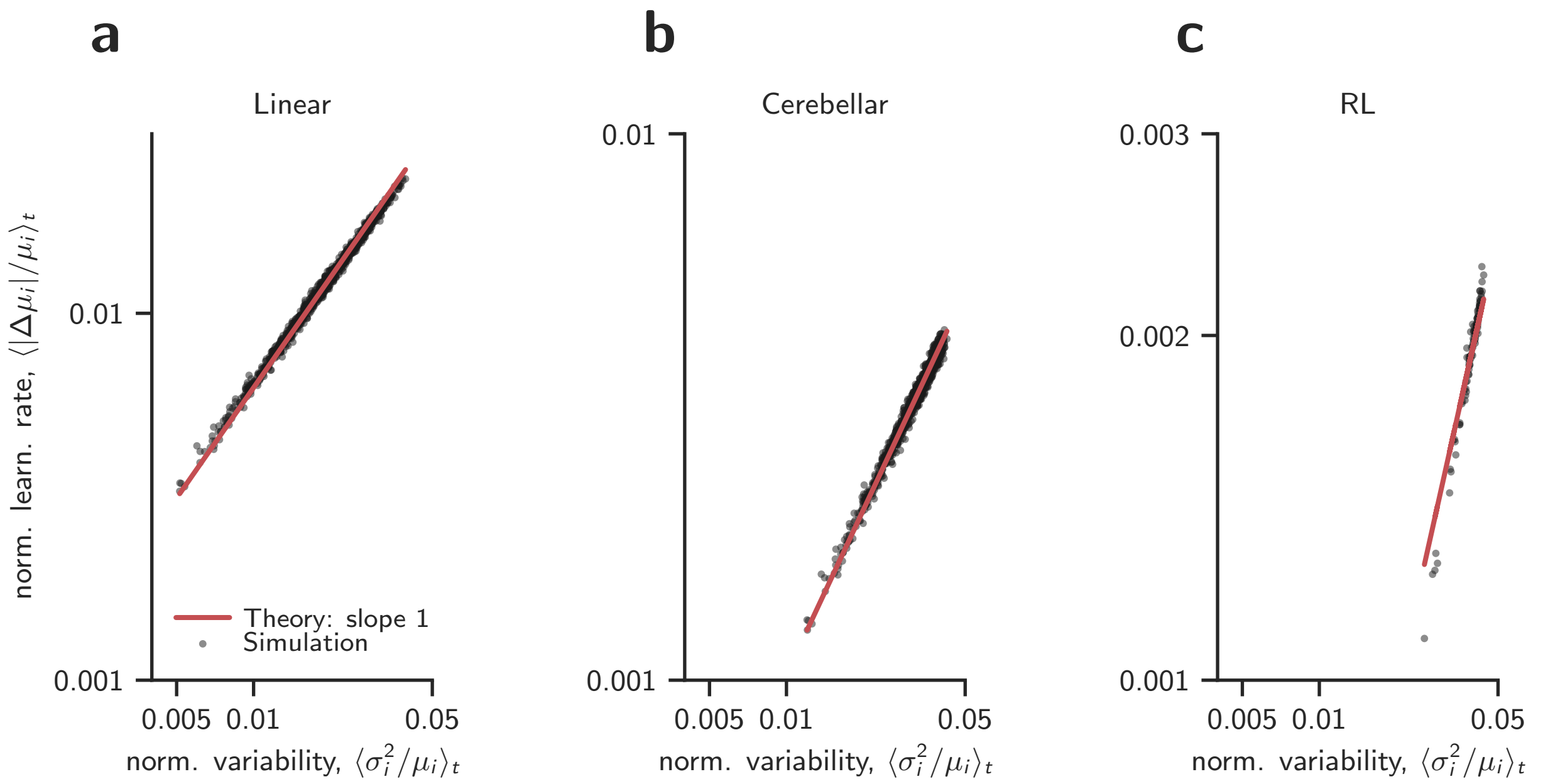

Figure S.7: Time average of normalized learning rate in our simulations, $\langle|\Delta\mu_i/\mu_i|\rangle_t$, versus normalized variability, $\langle\sigma_i^2/\mu_i\rangle_t$. As predicted in Sec. S3.3, $|\Delta\mu_i/\mu_i| \propto \sigma_i^2/\mu_i$, so the slope is 1 on a log-log plot. **a**. Linear feedback, $f_{\text{lin}} = \delta + \xi_\delta$. **b**. Cerebellar learning, $f_{\text{cb}} = \Theta(\delta + \xi_\delta - \theta)$. **c**. Reinforcement learning, $f_{\text{rl}} = -|\delta + \xi_\delta|$. Parameters from Table 1 (Sec. S5); $w_i$ was sampled from the posterior.

## S4 The relationship between variability and firing rate

To find the relationship between the mean and uncertainty, $\mu_i$ and $\sigma_i^2$, and the firing rate, $\nu_i$, we consider the steady state behavior of Eq. (S.68b), where $\langle\Delta s_i^2\rangle = 0$,

$$0 \approx \frac{x_i s_i^4 \mu_i^2}{\sigma_\delta^2} \frac{\sigma_\delta^2 - \text{Var}\left[f_{\text{lin}}\right]}{\sigma_\delta^2} + \frac{2}{\tau}\left(s_i^2 - s_{\text{prior}}^2\right). \tag{S.87}$$

Replacing $x_i$ by its average, $\nu_i \Delta t$, making the definition

$$\chi_i \equiv \frac{\sigma_\delta^2 - \text{Var}\left[f_{\text{lin}}\right]}{\sigma_\delta^2}, \tag{S.88}$$

solving for $s_i^2$, and then replacing $s_i^2$ using the approximate expression $s_i^2 \approx \sigma_i^2/\mu_i^2$ (Eq. (S.80)), we arrive at

$$\frac{\sigma_i^2}{\mu_i} \approx \frac{\left(2\mu_i^2 \tau \nu_i \Delta t \chi_i s_{\text{prior}}^2/\sigma_\delta^2 + 1\right)^{1/2} - 1}{\mu_i \tau \nu_i \Delta t \chi_i / \sigma_\delta^2}. \tag{S.89}$$

In the limit that the firing rate is sufficiently large,

$$\frac{\mu_i^2 \tau \nu_i \Delta t \chi_i s_{\text{prior}}^2}{\sigma_\delta^2} \gg 1, \tag{S.90}$$

Eq. (S.89) simplifies considerably,

$$\frac{\sigma_i^2}{\mu_i} \approx \frac{\sigma_\delta}{\sqrt{\nu_i \Delta t}} \left( \frac{2 s_{\text{prior}}^2}{\tau \chi_i} \right)^{1/2} . \tag{S.91}$$

This last expression tells us that for sufficiently large presynaptic firing rate, the normalized variability scales as $\nu^{-1/2}$. However, for small firing rate there are nontrivial corrections. Using Eq. (S.89), and noting that $\mu_i$ is independent of the presynaptic firing rate, $\nu_i$, our analysis predicts that the normalized variability should depend on presynaptic firing rate as

$$\frac{\sigma_i^2}{\mu_i} = a \, \frac{\sqrt{\nu_i/\nu_0 + 1} - 1}{\nu_i} . \tag{S.92}$$

In Fig. S.8 (top row) we plot $\sigma_i^2/\mu_i$ versus firing rate on a log-log plot, along with the prediction given in Eq. (S.92) (with parameters $a$ and $\nu_0$ chosen to minimized the mean squared error between the data and the prediction). The fit is remarkably good, especially given the rather gross approximations that we made. Note that the correction to the $1/\sqrt{\nu}$ scaling is most pronounced for reinforcement learning. In hindsight, that's expected: $\chi$ is the difference between the variance of $\delta$ and the variance of $f_{\text{lin}}$; for reinforcement learning, the goal of the learning rule is to make these two quantities as close as possible (see Methods, Eq. (M.47), and note that $f_{\text{rl}}^2 = f_{\text{lin}}^2$), so $\chi$ is small; and small $\chi$ implies large $\nu_0$, and thus a large correction.

Because the normalized learning rate, $|\Delta\mu_i/\mu_i|$, is proportional to $\sigma_i^2/\mu_i$ (see Fig. S.7), we expect $|\Delta\mu_i/\mu_i|$ to also be well fit by Eq. (S.92), with approximately the same values of $a$ and $\nu_0$ as in the top row of Fig. S.8. We do indeed see this in our simulations (Fig. S.8, bottom row).

## S5 Simulations details

Our simulations were relatively straightforward: either we iterated the single neuron update rules (Eqs. (M.41), (M.43) or (M.47) for the Bayesian rules and Eqs. (M.42), (M.46) or (M.48) for the classical ones) or solved the differential equation for the recurrent neural network (Methods, Eq. (M.60) for the Bayesian rules or the same equation but with $\eta_i$ set to a constant, independent of time or synapse, for the classical ones).

For Figs. 2 and 3 (single neuron update rules; see Table 1), we did not sample for the linear and cerebellar rules, and we used sampling with variance proportional to the mean for the reinforcement learning rule, as described in Methods, Sec. M1.2. For both figures, we ran the simulations for 500 OU time constants. In Fig. 2, we plotted the last three OU time constants. For Fig. 3, we used the first two OU time constants for burn-in and then computed the mean squared error using the remaining 498 OU time constants.

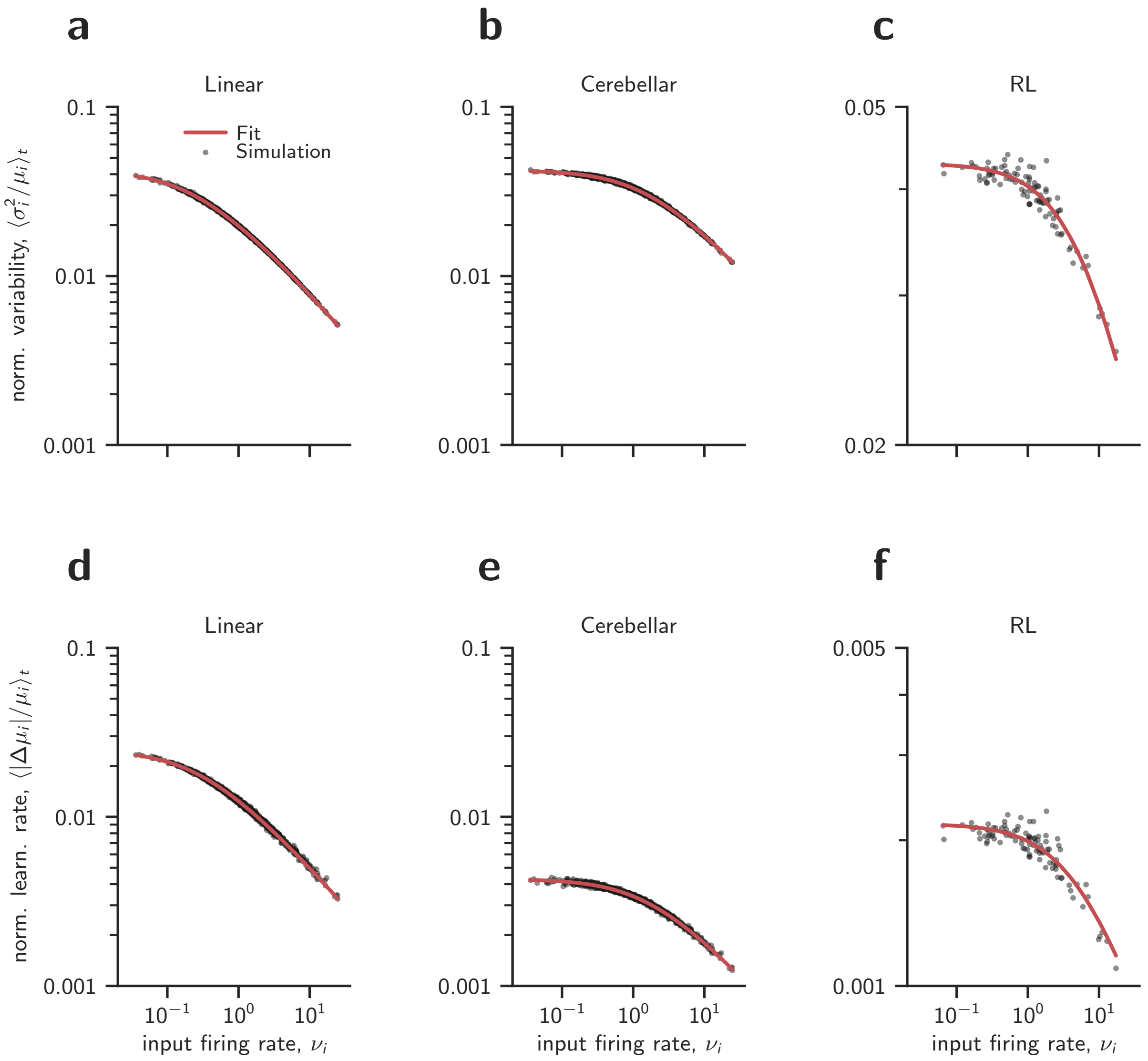

Figure S.8: Time average of normalized variability in our simulations, $\langle \sigma_i^2/\mu_i \rangle_t$ (top row), and normalized learning rate, $\langle |\Delta\mu_i/\mu_i| \rangle_t$ (bottom row), versus presynaptic firing rate, $\nu_i$. Red lines are best fit (minimum mean squared error) to Eq. (S.92). As predicted in that equation, both have a slope of $-1$ for sufficiently high firing rate. **a** and **d**. Linear feedback, $f = \delta + \xi_\delta$. **b** and **e**. Cerebellar learning, $f = \Theta\left(\delta + \xi_\delta - \theta\right)$. **c** and **f**. Reinforcement learning, $f = -|\delta + \xi_\delta|$. Parameters from Table 1 (Sec. S5); $w_i$ was sampled from the posterior.

The parameters of the recurrent neural network (Methods, Eq. (M.49)) are given in Table 2. Those parameters mainly describe the weights and target functions, which were chosen

| Parameter | Value | Basis |
|---|---|---|
| $m_{\text{prior}}$ | -0.669 | Matched to data from [2] (Sec. S6.1) |
| $s^2_{\text{prior}}$ | 0.863 | Matched to data from [2] (Sec. S6.1) |
| $n$ (linear, cerebellar) | 1000 | Number of synapses; offers a good trade-off between biological realism [3] and computational tractability |
| $n$ (reinforcement) | 100 | Number of synapses; uses a reduced number because of the difficulty of the learning problem |
| $\tau$ (linear, cerebellar) | $10^5$ | corresponds to 1,000 s (Sec. S6.3) |
| $\tau$ (reinforcement) | $5 \times 10^5$ | corresponds to 5,000 s (Sec. S6.3) |
| $\Delta t$ | 10 ms | Time step; set to the typical membrane time constant [4] |
| $\sigma_0$ | 2 mV | Standard deviation of combined membrane potential and feedback signal noise. |
| $k$ | 0.0877 | Normalized variability, matched to data [2] (Sec. S6.2) |
| $\theta$ | -4.2 | Used for the cerebellar feedback; chosen so that $f_{\text{cb}}$ is 1 about every 100 time steps, corresponding to a feedback signal of about 1 Hz. |

Table 1: Parameters used in the single neuron simulations (Figs. 2 and 3, main text).

as follows. The recurrent weights were sampled randomly as in [5],

$$J_{ij} = \xi_{ij} J^0_{ij} \tag{S.93}$$

where $\xi_{ij}$ is a Bernoulli random variable with probability $p$, and $J^0_{ij}$ is Gaussian and independent,

$$\xi_{ij} \sim \text{Bernoulli}(p) \tag{S.94a}$$

$$J^0_{ij} \sim \mathcal{N}\left(0, \frac{g^2}{pN}\right) \tag{S.94b}$$

(recall that $N$ is the number of neurons in the network). The feedback weights, $A_i$, were also random, and chosen to be uniform between $-1$ and 1. The readout weights, $w_i$, were initialized to be Gaussian and independent,

$$w_i \sim \mathcal{N}\left(0, \frac{1}{N}\right) . \tag{S.95}$$

The inputs, $I_i$, were transient pulses of 10 ms duration with amplitude sampled from a uniform distribution between $-1$ and 1. The initial conditions, $x_i(0)$, were obtained by first running the network with no input ($I_i = 0$) for 2000 ms. This allowed the network to relax to a strange attractor determined by its intrinsic dynamics.

| Parameter | Value | Basis |
|---|---|---|
| $N$ | 500 | Number of neurons in the network |
| $g$ | 1.5 | Scale factor for weights; this put the network in a mildly chaotic regime |
| $p$ | 0.1 | Connection probability |
| $\tau_m$ | 10 ms | Time constant of the $v_i$ |
| $\tau_{\text{target}}$ | 500 ms | Period of sampled functions |

Table 2: Parameters used in the recurrent neural network simulations (Fig. 4)

The target functions, $V_{\text{tar}}(t)$, were sampled randomly from a Gaussian process with periodic kernel

$$k(t, t') = \exp\left[-2\sin^2\left(\frac{\pi|t - t'|}{\tau_{\text{target}}}\right)\right] . \tag{S.96}$$

This kernel ensures that the sampled functions are periodic with period $\tau_{\text{target}}$.

We sampled 20 random target functions and 20 random recurrent networks as indicated above. For each network/target pair, we continually trained for 6000 periods of the target function (we show only the first 1000 in Fig. 4 because the Bayesian learning rules converged by then). Every 100 periods we stopped training, ran the network for 50 periods under the current readout weights (without feedback), and computed the mean squared error (MSE) between the readout and the true target function.

All figures show median MSE over all 400 network/target pairs. Error bars are bootstrapped confidence intervals for the median, using the percentile bootstrap method. Median was used rather than mean because there were always a few pairs in which learning failed. In these cases, the MSE remained very high, thus distorting the mean MSE in a way that did not allow for an illustrative comparison.

# S6 Choosing parameters

Below we summarize how we chose parameters for the single neuron update rules.

## S6.1 Estimate of priors from data

To determine the prior mean and variance of the log weights, $m_{\text{prior}}$ and $s^2_{\text{prior}}$, we used connectivity data from Ref. [2] (http://plasticity.muhc.mcgill.ca/DataPage/DataPage.html). The data is the mean and variance (over trials) of synaptic strength from paired recordings in rat visual cortex *in vitro*. To translate these to the mean and variance of the log-normal, we inverted Methods, Eq. (M.13),

$$m_i = \log \mu_i - \frac{1}{2} \log\left(1 + \frac{\sigma_i^2}{\mu_i^2}\right) \tag{S.97}$$

where $\mu_i$ and $\sigma_i^2$ are the mean and variance of the $i^{\text{th}}$ connection strength. The mean and variance of the prior are then given by the empirical mean and variance of the connections strengths,

$$m_{\text{prior}} = \frac{1}{M} \sum_{i=1}^{M} m_i \tag{S.98a}$$

$$s^2_{\text{prior}} = \frac{1}{M} \sum_{i=1}^{M} (m_i - m_{\text{prior}})^2 \tag{S.98b}$$

where $M$ (=852) is the number of connection strengths in the dataset. We are assuming that $\mu_i$ and $\sigma_i^2$ are good estimates of the true mean and variance of the log weights, and that an average over neurons is a good proxy for an average over time.

## S6.2 Variance versus mean: computing $k$

As discussed in Methods, Sec. M1.2, to make comparison with the classical reinforcement learning rule, we sample from the weights (see Eq. (M.14)). To do that, we assume that the variance is linear in the mean, and compute the slope, $k$, using, as above, data from Ref. [2] (http://plasticity.muhc.mcgill.ca/DataPage/DataPage.html). A plot of the variance, $\sigma^2$, versus the mean, $\mu$, is given in Fig. S.9, along with the best fitting line, $\sigma^2 = 0.0877\,\mu$.

## S6.3 Timescale for weight drift

As shown in Sec. S4, the normalized variability, $\sigma_i^2/\mu_i$, is related to the timescale for weight drift, $\tau$. To get a very approximate estimate for $\tau$, we replace all quantities in Eq. (S.91) by

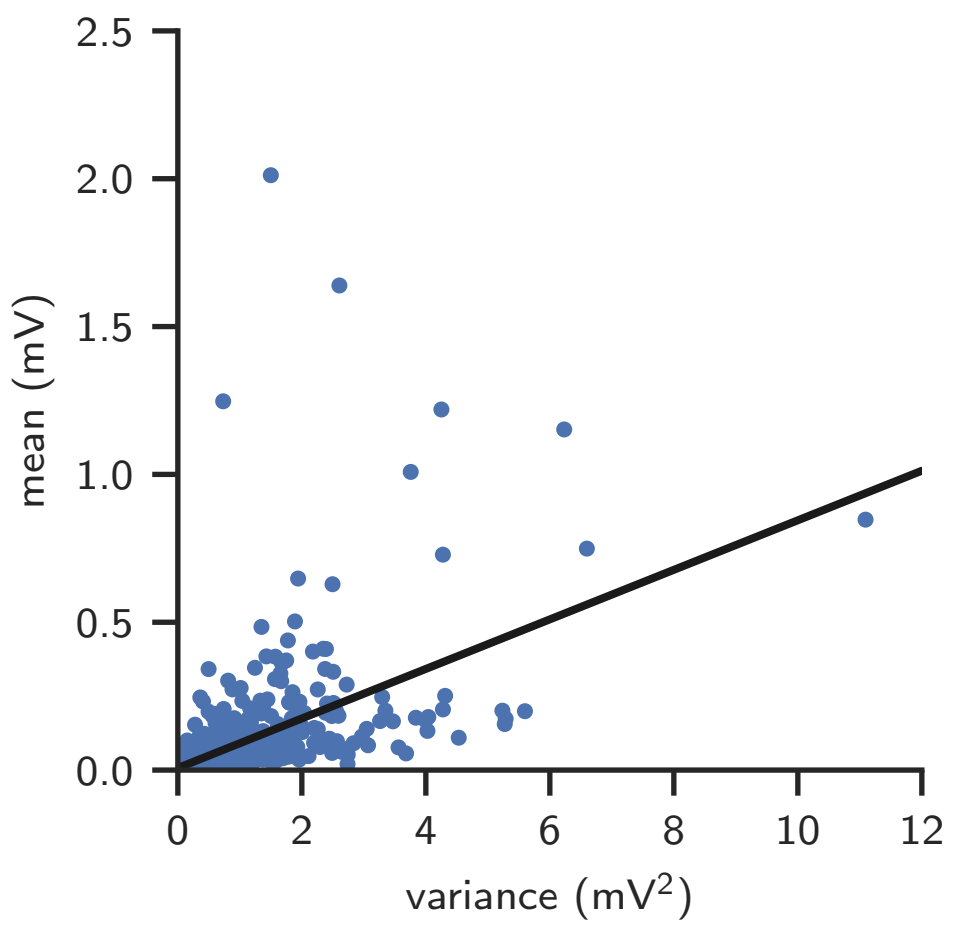

Figure S.9: Variance versus mean, using data (blue dots) taken from Ref. [2]. A least squares fit to the data gives $\sigma^2 = k\mu$ with $k = 0.0877$ (black line).

an average over neurons, for which we use an overline. Doing this, and solving that for $\tau$, we have

$$\tau \approx \frac{\sigma_\delta^2}{\overline{\nu}\Delta t} \frac{2s_{\text{prior}}^2}{\overline{\chi}} \frac{1}{\left(\overline{\sigma^2/\mu}\right)^2} . \tag{S.99}$$

To determine the size of $\sigma_\delta^2$, we use Methods, Eq. (M.38), assume sampling (Eq. (M.39)) and small $\nu_i \Delta t$, and ignore $\sigma_0^2$; that gives us

$$\sigma_\delta^2 \approx 2n\sigma_{\text{prior}}^2 \overline{\nu}\Delta t \tag{S.100}$$

where, recall, $n$ is the number of presynaptic neurons. For the remaining parameters we use $\overline{\sigma^2/\mu} = k = 0.0877$ and $\sigma_{\text{prior}}^2 = m_{\text{prior}}^2[e^{s_{\text{prior}}^2} - 1]$ (Methods, Eq. (M.13)). The latter quantity is approximately equal to 0.61 (Table 1), so Eq. (S.99) becomes

$$\tau \approx \frac{270\, n}{\overline{\chi}} . \tag{S.101}$$

For linear and cerebellar learning, $n = 1000$ and $\overline{\chi}$ is $\mathcal{O}(1)$; consequently, $\tau$ should be on the order of $3 \times 10^5$; we used $10^5$ in our simulations. (Recall that $\tau$ is measured in timesteps, which are 10 ms, so $\tau = 10^5$ corresponds to 1000 s.) For reinforcement learning, we reduce $n$ by a factor of 10, to 100 (because learning is hard). However, as discussed in Sec. S4, $\overline{\chi}$ is small; we therefore used $\tau = 5 \times 10^5$. While the approximations are somewhat crude, in simulations (Fig. S.8, top row) the normalized variability is indeed near its experimental value, 0.0877.

### S6.4 Firing rate data used for Fig. 5

To obtain the $p$-value for Fig. 5, we performed standard linear regression: we regressed log(variance/mean) against log(firing rate) and log(mean); the former to test our prediction and the latter to eliminate the PSP amplitude as a possible confound. To estimate the firing rate, we took the mean of a FOOPSI-based firing rate estimate [6] supplied to us by the authors of [7]. This estimate is proportional to the true firing rate [8]; because our predicted relationship was linear on a log-log plot, the constant of proportionality plays no role. Using this approach, the best fit line was statistically significantly different from zero ($p < 0.003$, $t$-test, $n = 135$), and its slope, $-0.62$ was not significantly different from our prediction, $-1/2$ ($p = 0.57$, $t$-test).

## S7 Synaptic Sampling

Here we provide an expanded normative argument for Synaptic Sampling. The argument starts with the observation that to select the correct action, knowing the uncertainty in task relevant quantities is critical [9]. For instance, to decide whether you can jump over a puddle without getting your feet wet, you need more than just an estimate of mean landing location; you also need an estimate of uncertainty (Fig. S.10). Uncertainty about the landing location comes from two sources: uncertainty about the current state of the world and uncertainty about the target weights (i.e., the weights that would give the best estimate of landing location). To see how the brain might compute uncertainty in landing location, consider a simplified scenario in which we use $\mathbf{x}_{\text{tar}}$ to denote the best possible spike-based representation of the true state of the external world. The neuron's estimate of landing location is a function of the neuron's output, $V$, so the optimal estimate of landing location is given by the target output,

$$V_{\text{tar}} = \mathbf{w}_{\text{tar}} \cdot \mathbf{x}_{\text{tar}}. \tag{S.102}$$

The assumption that the synapse combines $\mathbf{w}_{\text{tar}}$ and $\mathbf{x}_{\text{tar}}$ via a dot product is for simplicity only; the cell could use any nonlinear relationship and our arguments would hold.

Of course, the brain knows neither the target weights, $\mathbf{w}_{\text{tar}}$, nor the true state of the external world, $\mathbf{x}_{\text{tar}}$. The brain could compute a "best guess" of $\mathbf{x}_{\text{tar}}$, and the neuron could use a "best guess" of $\mathbf{w}_{\text{tar}}$, resulting in

$$V_{\text{best guess}} = \mathbf{w}_{\text{best guess}} \cdot \mathbf{x}_{\text{best guess}} \,. \tag{S.103}$$

However, this scheme is unable to give an estimate of uncertainty — so offers little guidance as to whether or not to jump over the puddle.

To get an estimate of uncertainty, it is necessary to account for uncertainty both in the state of the world, $\mathbf{x}_{\text{tar}}$, and in the relationship between the state of the world and jump distance, parameterized by $\mathbf{w}_{\text{tar}}$. As information about $\mathbf{x}_{\text{tar}}$ comes from sensory data, and information about $\mathbf{w}_{\text{tar}}$ comes from training data (e.g., from past jumps), we can represent our (probabilistic) knowledge about these quantities as two distributions, $P(\mathbf{x}_{\text{tar}}|\text{Sensory Data})$ and $P(\mathbf{w}_{\text{tar}}|\text{Training Data})$. To combine these distributions into a distribution over $V_{\text{tar}}$, we need to integrate over all possible settings of $\mathbf{x}_{\text{tar}}$ and $\mathbf{w}_{\text{tar}}$,

$$P(V_{\text{tar}}|\text{Sensory Data}, \text{Training Data}) = \int d\mathbf{w}_{\text{tar}}\, d\mathbf{x}_{\text{tar}}\, P(V_{\text{tar}}|\mathbf{x}_{\text{tar}}, \mathbf{w}_{\text{tar}})\, P(\mathbf{x}_{\text{tar}}|\text{Sensory Data})\, P(\mathbf{w}_{\text{tar}}|\text{Training Data})\,. \tag{S.104}$$

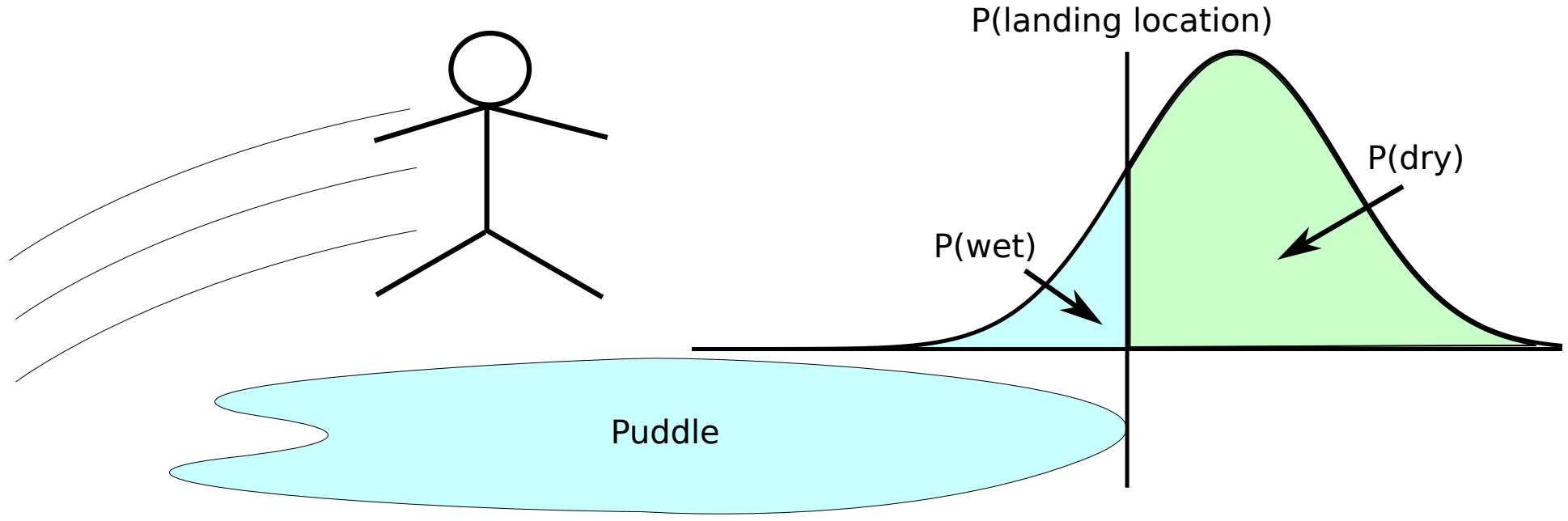

Figure S.10: A schematic diagram of a stick-person jumping over a puddle. The probability of landing in the puddle, $P(\text{wet})$, depends not only on the mean estimate, but also on the uncertainty.

It is difficult for neurons to compute this integral, as it is high dimensional and rarely has a closed form expression. However, by combining neural and synaptic sampling, it is possible for neural circuits to evaluate the integral via sampling; that is, by drawing samples, $V$, from the distribution,

$$V \sim P(V_{\text{tar}}|\text{Sensory Data}, \text{Training Data})\,. \tag{S.105}$$

To do that, we simply need to draw neural activity, $\mathbf{x}$, from its distribution given sensory data,

$$\mathbf{x} \sim P(\mathbf{x}_{\text{tar}}|\text{Sensory Data}) \tag{S.106}$$

(this is known as the neural sampling hypothesis [10, 11, 12]), and draw synaptic weights, $\mathbf{w}$, from their distribution given training data,

$$\mathbf{w} \sim P(\mathbf{w}_{\text{tar}}|\text{Training Data}) \tag{S.107}$$

(this is our Synaptic Sampling hypothesis). A sample of landing location is given by combining the sampled neural activity with the sampled weights, which could be done by a single neuron,

$$V = \mathbf{w} \cdot \mathbf{x}\,. \tag{S.108}$$

Thus, simply by drawing repeated samples, a single neuron can estimate uncertainty about $V$, and thus about landing location.

Our argument appears to assume that the brain uses the output of a single neuron to make predictions. This is not too implausible — the cerebellum does contain a large number of Purkinje cells [13] that are believed to use supervised learning to, among other things, make predictions. However, it is certainly possible that such a computation is performed by a large multi-layer network. As long as the network is effectively feedforward, we can still, by the logic described above, estimate its uncertainty by combining synaptic sampling with neural sampling.

## S8 Robustness

Our Bayesian update rules depend on a number of parameters, and in our simulations so far we set them to their theoretically optimal values. However, the brain can't do this; there will always be some model mismatch. Here we explore the robustness of Bayesian plasticity to this mismatch. For linear, cerebellar and reinforcement learning, there are three main parameters: $\sigma_0^2$, the combined noise in the feedback signal and membrane potential, and $m_{\text{prior}}$ and $s_{\text{prior}}^2$, the mean and variance that govern the random drift in synaptic weights (see Eqs. (M.41), (M.43) and (M.47)). For these parameters, we examine performance when they change by a factor of 2 in either direction relative to their nominal values (with, of course, the update rules assuming they have not changed). The results are shown in Figs. S.11a-c. The linear and cerebellar rules were robust with respect to changes in all these parameters. Sensitivity to $m_{\text{prior}}$ was highest, although $\pm 50\%$ changes in that parameter had little effect on the mean squared error. For the other two parameters, the mean squared error changed very little over the range tested. The reinforcement learning rule was more sensitive to parameters, especially $\sigma_0^2$, where small decreases led to an instability that greatly increased the mean squared error. However, increases had little effect. Overall, the Bayesian learning rule is relatively robust. It certainly does not require fine tuning.

We also looked at the effect of binary, rather than Gaussian, drift. That is, in Methods, Eq. (M.11), rather than drawing $\xi_{\text{tar},i}$ from a Gaussian distribution, we used $\xi_{\text{tar},i} = \pm 1$, with $+1$ and $-1$ occurring with equal probability. As can be seen in Fig. S.11d, the results were virtually identical to the Gaussian case. Given the small time step relative to the OU time constant, this is not surprising.

For the recurrent neural network, the main parameter in the update rule is the initial learning rate, $\eta_i(0)$ (Methods, see Eq. (M.60)). Performance is extremely insensitive to this parameter: it can change by a factor of 30 without affecting the mean squared error (Fig. S.12a). We also checked sensitivity to the parameter $g$, which determines how chaotic the network is (see Eq. (S.94b)). For values of $g$ that yielded low mean squared errors (1.25 and 1.5; we used 1.5 in our simulations), the Bayesian learning rules remained about an order of magnitude better than the classical ones (Fig. S.12b). Thus, Bayesian learning in the recurrent neural network is extremely robust.

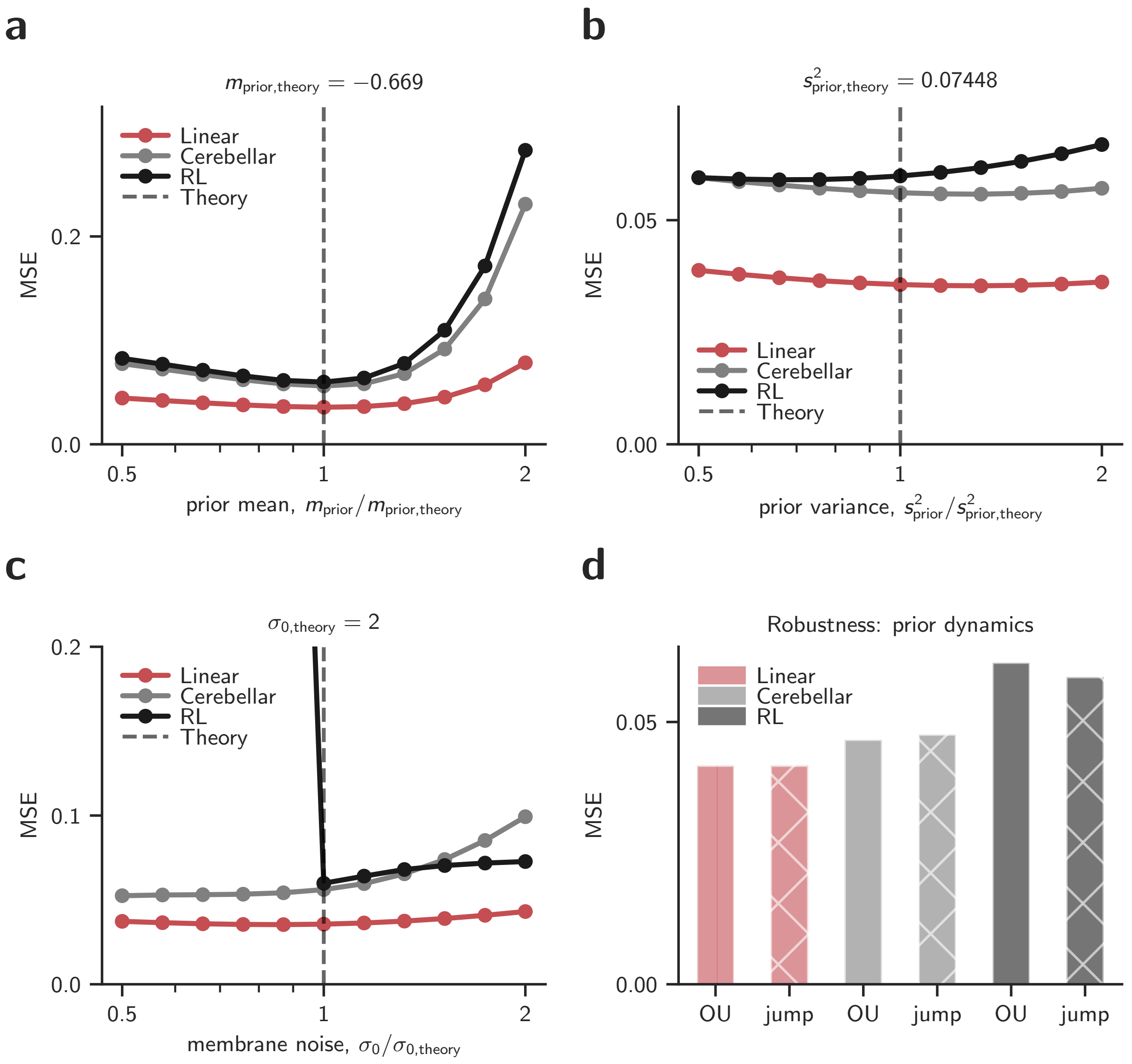

Figure S.11: Robustness with respect to parameters for the single neuron learning rules. In panels a-c, quantities with a subscript "theory" are the optimal values as predicted by our theoretical model. **a**. $m_{\text{prior}}$; mean of the Ornstein-Uhlenbeck (OU) process (Methods, Eq. (M.11)). **b**. $s^2_{\text{prior}}$; variance of the OU process (Eq. (M.11)). **c**. $\sigma_0$; standard deviation of the combined noise in the membrane potential and the feedback signal (Methods, Eq. (M.33)). **d**. Effect of using jump, rather than Gaussian, noise in the weight drift. For "OU", $\xi_{\text{tar},i} \sim \mathcal{N}(0, 1)$; for "jump", $\xi_{\text{tar},i}$ is $+1$ or $-1$, with equal probability.

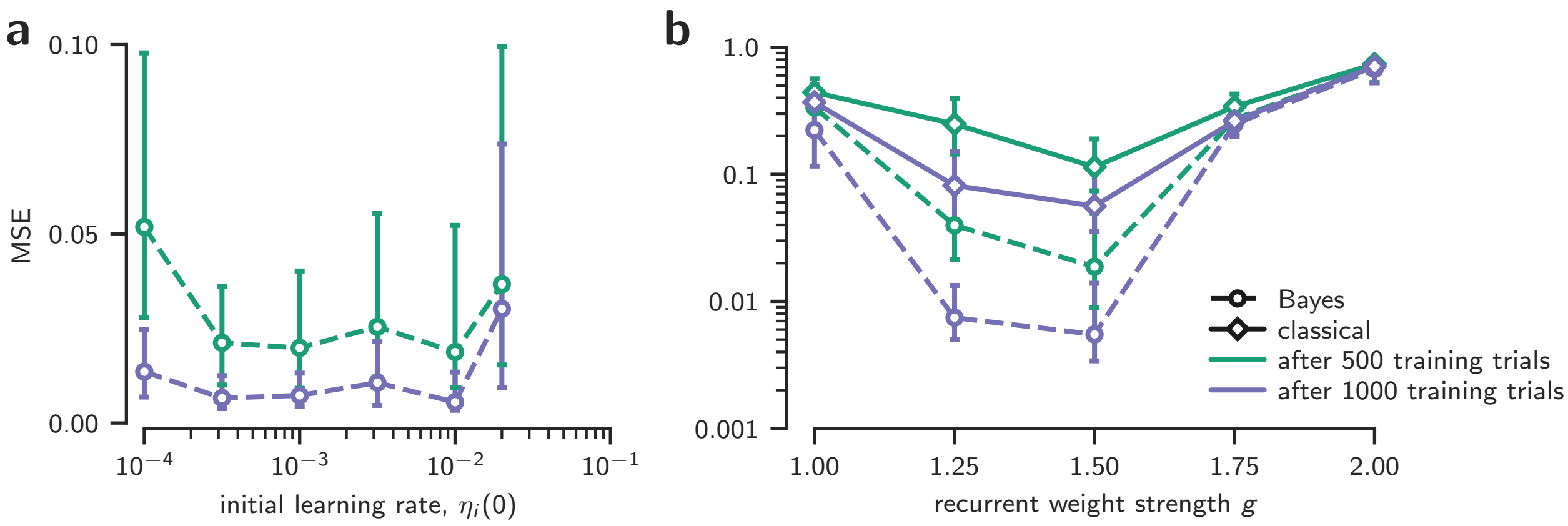

Figure S.12: Robustness with respect to parameters for the recurrent neural network. **a**. Mean squared error versus the initial learning rate, $\eta_i(0)$. The value used in our simulations was $\eta_i(0) = 10^{-2}$. **b**. Mean squared error of the Bayesian and classical learning rule as a function of $g$, the parameter that controls how chaotic the network is (see Eq. (S.94b); we used $g = 1.5$ in our simulations). In both panels we report median over $n = 400$ network/target pairs, and error bars are 95% confidence intervals computed using the percentile bootstrap.

# References

[1] A. Gelb. *Applied Optimal Estimation*. MIT Press, 1974.

[2] S. Song et al. “Highly nonrandom features of synaptic connectivity in local cortical circuits”. In: *PLoS Biology* 3.3 (2005), e68.

[3] T. Binzegger, R. Douglas, and K. Martin. “A quantitative map of the circuit of cat primary visual cortex”. In: *J. Neurosci.* 24 (2004), pp. 8441–8453.

[4] S. J. Tripathy et al. “Brain-wide analysis of electrophysiological diversity yields novel categorization of mammalian neuron types”. In: *Journal of Neurophysiology* 113.10 (2015), pp. 3474–3489.

[5] B. DePasquale et al. “Full-FORCE: A Target-Based Method for Training Recurrent Networks”. In: *PLoS ONE* 13.2 (2018), e0191527.

[6] J. T. Vogelstein et al. “Fast nonnegative deconvolution for spike train inference from population calcium imaging”. In: *Journal of Neurophysiology* 104.6 (2010), pp. 3691–3704.

[7] H. Ko et al. “The emergence of functional microcircuits in visual cortex”. In: *Nature* 496.7443 (2013), pp. 96–100.

[8] A. M. Packer et al. “Simultaneous all-optical manipulation and recording of neural circuit activity with cellular resolution in vivo”. In: *Nature Methods* 12.2 (2015), pp. 140–146.

[9] M. O. Ernst and M. S. Banks. “Humans integrate visual and haptic information in a statistically optimal fashion”. In: *Nature* 415.6870 (2002), pp. 429–433.

[10] P. O. Hoyer and A. Hyvarinen. “Interpreting neural response variability as Monte Carlo sampling of the posterior”. In: *Advances in Neural Information Processing Systems*. 2003, pp. 293–300.

[11] J. Fiser et al. “Statistically optimal perception and learning: from behavior to neural representations”. In: *Trends in Cognitive Sciences* 14.3 (2010), pp. 119–130.

[12] P. Berkes et al. “Spontaneous Cortical Activity Reveals Hallmarks of an Optimal Internal Model of the Environment”. In: *Science* 331.6013 (2011), pp. 83–87.

[13] P. Dean et al. “The cerebellar microcircuit as an adaptive filter: experimental and computational evidence”. In: *Nature Reviews Neuroscience* 11.1 (2010), pp. 30–43.